\documentclass{aa}
\usepackage[varg]{txfonts}
\usepackage{amssymb}
\usepackage{multirow}
\usepackage{subcaption}
\usepackage{natbib}
\usepackage{float}
\usepackage{caption}
\usepackage{hyperref}
\usepackage{threeparttable}

\hypersetup{
    colorlinks=true,
    linkcolor=blue,
    filecolor=magenta,      
    urlcolor=cyan,
    citecolor=blue,
    }

\bibpunct{(}{)}{;}{a}{}{,}

\begin{document}

\title{Spectral properties of ablating meteorite samples for improved meteoroid composition diagnostics}

\author{Pavol Matlovič\inst{1}
  \and Adriana Pisarčíková\inst{1}
  \and Veronika Pazderová\inst{1}
  \and Stefan Loehle\inst{2}
  \and Juraj Tóth\inst{1}
  \and Ludovic Ferri\`ere\inst{3,}\inst{4}
  \and Peter Čermák \inst{1}
  \and David Leiser\inst{2}
  \and Jérémie Vaubaillon\inst{5}
  \and Ranjith Ravichandran\inst{2,}\inst{6}
  }

\institute{Faculty of Mathematics, Physics and Informatics,
  Comenius University Bratislava, Mlynská dolina, 84248 Bratislava, Slovakia\\
  \email{matlovic@fmph.uniba.sk}  
  \and High Enthalpy Flow Diagnostics Group, Institute of Space Systems, University of Stuttgart, Pfaffenwaldring 29, 70569 Stuttgart, Germany
  \and Natural History Museum Vienna, Burgring 7, 1010 Vienna, Austria
  \and Natural History Museum Abu Dhabi, Jacques Chirac street, Al Saadiyat Island - Cultural District, Abu Dhabi, United Arab Emirates
  \and IMCCE, Observatoire de Paris, PSL, 77 Av Denfert Rochereau, 75014 Paris, France
  \and Institute of Advanced Engineering and Space Sciences, University of Southern Queensland, Australia
  }

\date{Accepted for publication 08/2024}

\abstract{Emission spectra and diagnostic spectral features of a diverse range of ablated meteorite samples with a known composition are presented. We aim to provide a reference spectral dataset to improve our abilities to classify meteoroid composition types from meteor spectra observations. The data were obtained by ablating meteorite samples in high-enthalpy plasma wind tunnel facilities recreating conditions characteristic of low-speed meteors. Near-UV to visible-range (320 -- 800 nm) emission spectra of 22 diverse meteorites captured by a high-resolution Echelle spectrometer were analyzed to identify the characteristic spectral features of individual meteorite groups. The same dataset captured by a lower-resolution meteor spectrograph was applied to compare the meteorite data with meteor spectra observations. Spectral modeling revealed that the emitting meteorite plasma was characterized by temperatures of 3700 -- 4800 K, similar to the main temperature component of meteors. The studied line intensity variations were found to trace the differences in the original meteorite composition and thus can be used to constrain the individual meteorite classes. We demonstrate that meteorite composition types, including ordinary chondrites, carbonaceous chondrites, various achondrites, stony-iron and iron meteorites, can be spectrally distinguished by measuring relative line intensities of Mg I, Fe I, Na I, Cr I, Mn I, Si I, H I, CN, Ni I, and Li I. Additionally, we confirm the effect of the incomplete evaporation of refractory elements Al, Ti, and Ca, and the presence of minor species Co I, Cu I, and V I.}

\keywords{meteorites, meteors, meteoroids -- techniques: spectroscopic}
\maketitle

\section{Introduction} \label{sec:intro}

Most of our current knowledge about the composition of small interplanetary bodies comes from laboratory studies of meteorites. Yet, meteorite findings are rare and only provide data on a biased sample of bodies strong enough to withstand atmospheric entry. Additional important insights come from spectral surveys of asteroids and comets, studies of micrometeorites and interplanetary dust particles, or sample return missions. Still, the full compositional diversity and distribution of materials within the wide population of small Solar System bodies remain largely unexplored. Significant benefits yet to be fully utilized can be extracted from meteor observations. The Earth's atmosphere daily encounters 30 to 150 tons of interplanetary matter in the form of meteoroids, micrometeoroids, and dust \citep[][and references therein]{2017P&SS..143...21D}. This material contains valuable information about the composition of interplanetary bodies arriving from various regions of the Solar System. The ablation of meteoroids in our atmosphere enables their convenient detection and investigation through meteor observation techniques. The extent of meteor analyses is advancing globally, driven by the availability of optical technologies that enhance the quality and quantity of the collected data. Meteor trajectory measurements allow us to characterize the dynamical origin of these bodies and often link them to their parent asteroids and comets. However, spectral research within current global meteor networks is still somewhat limited.

Various photographic and video techniques have been deployed to survey spectral properties of meteors since the early 20th century \citep[overview in][and references therein]{1980IAUS...90..121M, 2015A&A...580A..67V}. In the last two decades, most of the progress in meteor spectral research was focused on low-resolution video data, which generally allow higher detection rates, but are often affected by line blending and saturation. The method of spectral classification \citep{2005Icar..174...15B} has been proposed to study spectral differences of mm-sized meteoroids based on the variations of Mg, Na, and Fe atomic emission. This method was since applied in multiple studies \citep[e.g.][]{2019A&A...621A..68V, 2020P&SS..19405040A, 2020P&SS..18004773R}, and also to survey populations of larger, cm-sized meteoroids \citep{2019A&A...629A..71M}. Different spectral types of meteors and the effects of space weathering on meteoroids have been recognized using this method. However, the ability to accurately classify different meteoroid composition types, in part consisting of the range of meteorite types found on Earth, has not yet been achieved. 

Here we present results from our project MetSpec focused on linking emission spectra of various meteorite samples ablated in a plasma wind tunnel with spectra acquired from meteor observations. Our main goal is to identify and quantify distinct spectral features of specific meteorite types with a known bulk composition, which can be used for improved composition diagnostics of meteoroids captured by meteor spectrographs. The wind-tunnel meteorite ablation experiments, which form the basis of the data analyzed here, were studied by numerous instruments, in part described in a recent special collection of papers \citep[see overview in][]{2024Icar..40715791T}. Here we focus on analyzing and interpreting emission spectra in the visible and near-UV range. Partial results from the spectral analysis mainly focused on the emission of H and CN from meteorite samples were already presented in \citet{2020Icar..34713817M, 2022MNRAS.513.3982M, 2023Icar..40415682P}.

Since the experiment conditions are representative of a real atmospheric flight of a low-velocity meteor \citep{2017ApJ...837..112L, 2024Icar..40715817L}, the reference ablated meteorite data used is considered analogous to asteroidal meteoroids observed during meteor observations. The emission conditions of real meteors can vary widely, mainly depending on their speed and composition. Therefore, the application and interpretation of some of the presented diagnostic spectral metrics in faster cometary meteors must be performed carefully. Modeling emission spectra at various plasma conditions can help connect the two datasets. On a large scale, this is beyond the scope of this work. We have, however, performed spectrum modeling assuming relevant conditions attained in the plasma wind tunnel to better interpret some of the observed features and comment on the expected changes in spectral properties at different plasma conditions relevant to meteor events. 

In Section \ref{sec:methods} we present the utilized instruments and methods. Section \ref{results} describes the analyzed diagnostic spectral features in different meteorites. The summary of identified characteristic spectral properties of individual meteorite types is given in Section \ref{charcteristics}, followed by the conclusions.

\section{Instrumentation and methods} \label{sec:methods}

To reference emission spectra of various meteorite types presented here were collected in wind tunnel ablation experiments with the High Enthalpy Flow Diagnostics Group (HEFDiG), Institute of Space Systems in Stuttgart (IRS), Germany. Details of the performed laboratory ablation experiments and the applied instrumentation were extensively described in \citet{2024Icar..40715791T}, \citet{2024Icar..40715817L}, \citet{2023Icar..40415682P}, and references therein. Here we summarize the main experiment conditions and instrumentation characteristics relevant to the presented spectral analysis. For more information, please refer to the abovementioned publications. 

The applied flow condition representative of a low-speed meteoroid entry in the IRS plasma wind tunnel was developed by \citet{2017ApJ...837..112L}. The meteorite ablation experiments were carried out at a flow condition characterized by a local mass-specific enthalpy of 70 MJ kg\textsuperscript{-1} at a stagnation pressure of $\sim$24\,hPa. This corresponds to the entry of a of $\sim$4\,cm meteoroid at an altitude of $\sim$80\,km in the Earth's atmosphere, with an assumed meteoroid entry speed of $\sim$12\,km s\textsuperscript{-1} \citep{2024Icar..40715817L}. In this sense, the collected data represent analogs of slow meteor events caused by an atmospheric entry of an asteroidal meteoroid. A total of 22 different ablated meteorite samples are analyzed in this work. The meteorite classification and bulk chemical composition of major elements are provided in Table \ref{T1}. A majority of the meteorite samples used for this work were obtained from the Natural History Museum Vienna, the rest were owned by Comenius University Bratislava or sourced from meteorite dealers. The obtained observations present the largest dataset of laboratory meteor analogs collected to date. 

The visible-range emission spectra of meteorites analyzed here were captured by two instruments. Our analysis is based mainly on the high-resolution data from HEFDiG's Echelle spectrometer LTB ARYELLE 150. It is a fiber-fed system operating in a 250–880 nm wavelength range, with spectral dispersion varying from 43 pm px\textsuperscript{-1} to 143 pm px\textsuperscript{-1} within this wavelength interval (resolving power up to R $\approx$ 11 000). The Echelle data were calibrated according to the procedure described in \citet{2024Icar..40715817L}. The sum of all spectrum frames from the meteorite ablation exposure was created as a representative spectral profile of each meteorite. The differential ablation of meteorites was also studied by analyzing their spectra from individual frames as a function of time.

Meteorite emission spectra were also captured by the meteor spectrograph AMOS-Spec-HR operated by the Comenius University, Bratislava within their global AMOS (All-sky Meteor Orbit System) meteor network \citep{2022MNRAS.513.3982M, 2023AdSpR..71.3249Z}. This system uses a 1000\,gr/mm holographic grating in front of a 6\,mm/F1.4 lens and a DMK~33UX252 digital camera with a pixel resolution of 2048x1536 at a frame rate of 20\,fps. The field of view is 60x45 degrees and the limiting stellar magnitude is +4. The spectrograph has a resolution R $\approx$ 550. The use of this astronomical spectrograph has allowed direct comparison between the laboratory meteor analogs and real meteor observations. An adjustable slit was mounted on the wind tunnel window to limit the light source. Since AMOS is a sensitive instrument designed for distant night-sky observations, the camera gain had to be adjusted during each experiment depending on the ablation brightness. Furthermore, due to the close distance of the ablating meteorite, the emission spectra included a strong continuum from the heated meteorite, which was often saturated. The emission spectra were therefore scanned and analyzed above and below this continuum, i.e. from the heated meteor plasma surrounding the meteorite. 

\begin{figure*}
\centerline{\includegraphics[width=.95\textwidth,angle=0]{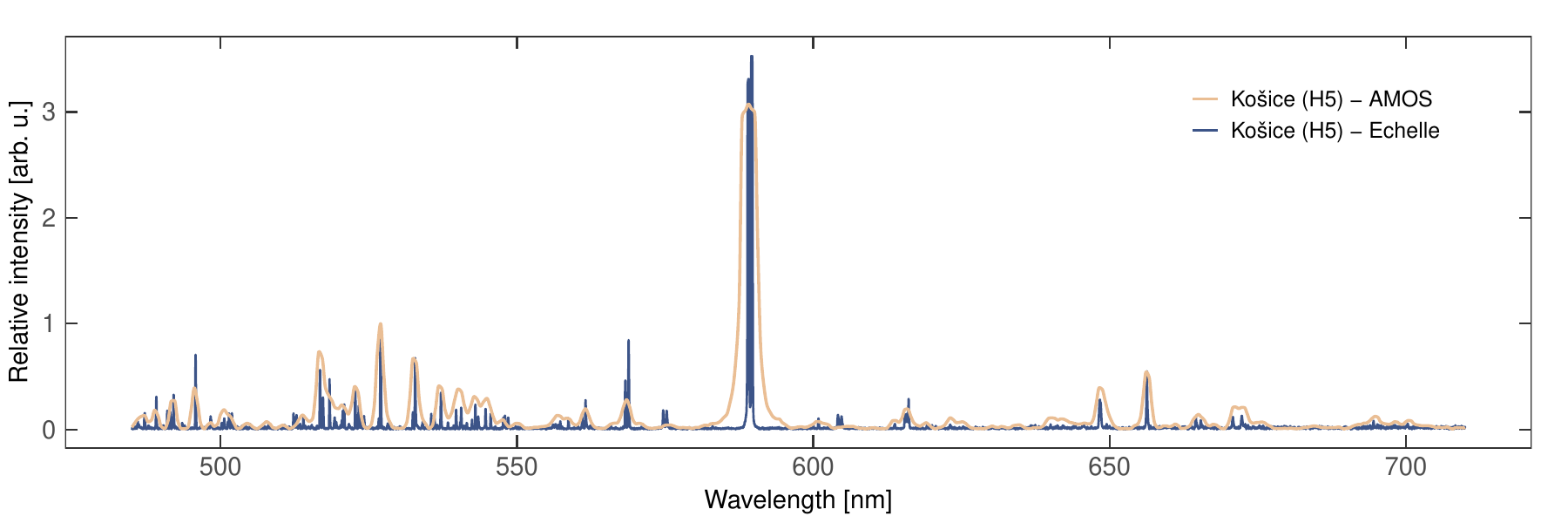}}
\caption[f1]{Emission spectrum of the H5 ordinary chondrite Košice. A comparison between the spectra captured by the Echelle (blue) and AMOS spectrographs (yellow) in the 480 - 700 nm range is displayed. The spectra are normalized to unity at 527.0 nm.}
\label{ech_vs_amos}
\end{figure*}

\begin{figure*}[]
\centerline{\includegraphics[width=\textwidth,angle=0]{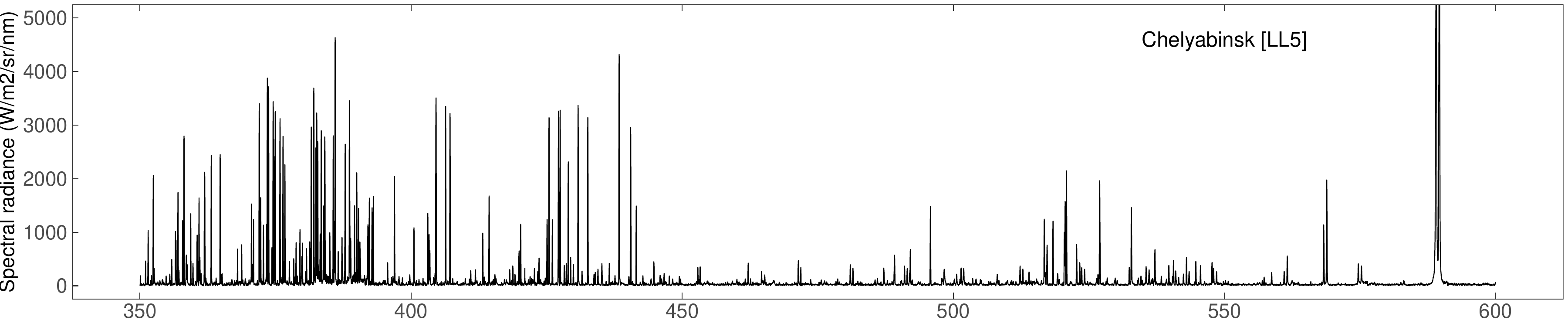}}
\centerline{\includegraphics[width=\textwidth,angle=0]{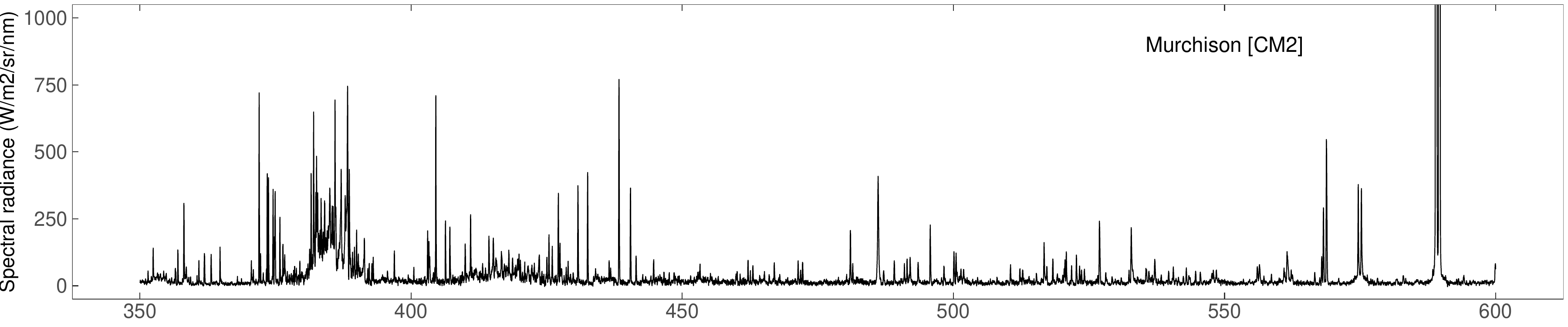}}
\centerline{\includegraphics[width=\textwidth,angle=0]{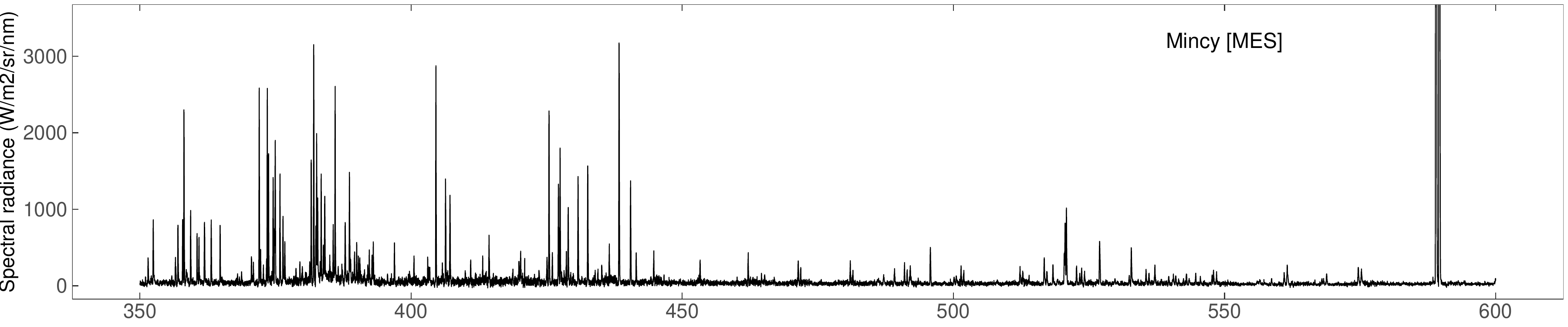}}
\centerline{\includegraphics[width=\textwidth,angle=0]{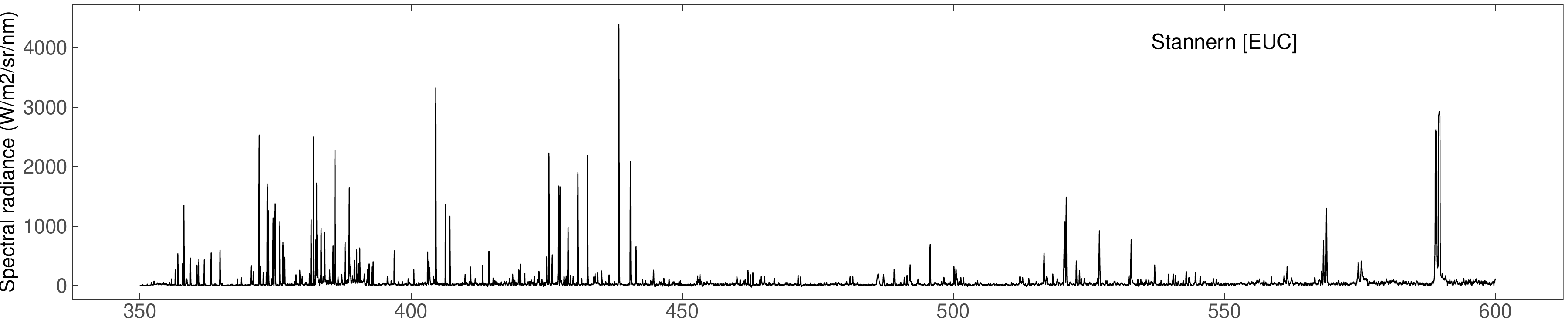}}
\centerline{\includegraphics[width=\textwidth,angle=0]{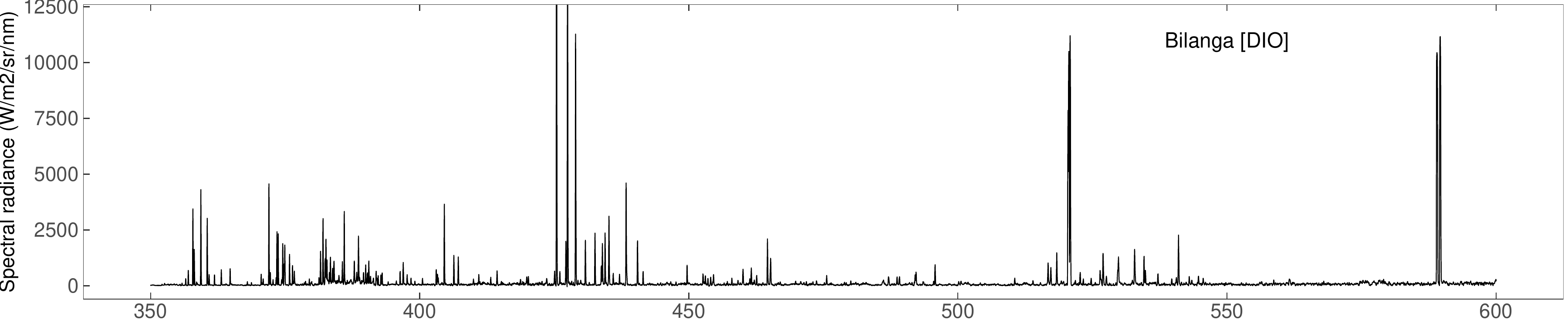}}
\centerline{\includegraphics[width=\textwidth,angle=0]{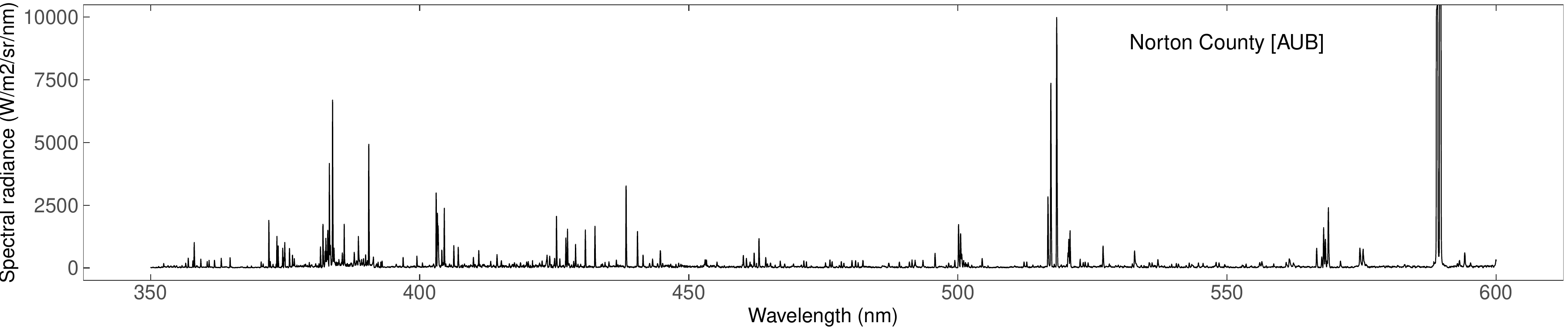}}
\caption[f1]{Examples of the emission spectra of ablating meteorites captured by the Echelle spectrometer. Only the wavelength range 350 -- 600 nm is displayed for better visibility. Full spectra were captured in the 320 -- 880 nm range.}
\label{profiles}
\end{figure*}

The Echelle and AMOS spectra of the ordinary chondrite Košice are compared in Fig. \ref{ech_vs_amos}. Similar spectral features and time-resolved behavior were observed in both obtained datasets. Due to the abovementioned limitations, the quality of the AMOS data was more sensitive to the individual experiment conditions and manual instrument adjustment. Since we aim to provide the best available reference spectral data of meteorites, we further focus on the high-resolution Echelle data. Examples of the emission spectra of meteorite samples captured by the Echelle spectrometer are displayed in Fig. \ref{profiles}. The reduced and calibrated spectral data were used to fit and measure integrated line intensities following the procedures described in \citet{2019A&A...629A..71M, 2022MNRAS.513.3982M}. For some applications, a radiative transfer model assuming local thermal equilibrium was used to evaluate the radiating plasma conditions using our custom fitting software and the NIST LIBS tool \citep{NISTtools}. It was found that for most experiments, the radiating plasma was close to optically thin, so self-absorption was not considered for the scope of this work. 

\begin{table*}
\small\begin{center}
\caption{Ablated meteorites analyzed in this work.}
\label{T1}
\resizebox{\textwidth}{!}{
\begin{tabular}{lllrrrrrrrrrrrrl}
\hline \\[-4pt]
\multicolumn{1}{l}{\textbf{Name}}& %
\multicolumn{1}{l}{\textbf{Type}}& %
\multicolumn{1}{l}{\textbf{Rec.}}& %
\multicolumn{1}{c}{\textbf{Fe}}& % 
\multicolumn{1}{c}{\textbf{Si}}& %
\multicolumn{1}{c}{\textbf{Mg}}& %
\multicolumn{1}{c}{\textbf{Ca}}& %
\multicolumn{1}{c}{\textbf{Al}}& %
\multicolumn{1}{c}{\textbf{Cr}}& %
\multicolumn{1}{c}{\textbf{Mn}}& %
\multicolumn{1}{c}{\textbf{Na}}& %
\multicolumn{1}{c}{\textbf{K}}& %
\multicolumn{1}{c}{\textbf{Ni}}& %
\multicolumn{1}{c}{\textbf{C}}& %
\multicolumn{1}{c}{\textbf{T [K]}}& %
\multicolumn{1}{c}{\textbf{Bulk ref.}} \\ 
\hline \\ [-2pt]
Košice          & H5 & fall & 28.85 & 16.51 & 13.80 & 1.14 & 1.07 & 0.350 & 0.230 & 0.600 & 0.090 & 1.00 & 0.05 & 4660 & [1] \\ [2pt]
Pultusk         & H5 & fall & 27.18 & 17.04 & 14.32 & 1.30 & 1.34 & 0.253 & 0.194 & 0.616 & 0.075 & 1.61 & - & 3770 & [2], [3] \\ [2pt]
Buzzard Coulee  & H4 & fall & - & - & - & - & - & - & - & - & - & - & - & 4020 & N/A \\ [2pt]
NWA 869         & L3-6 & find & 19.47 & 18.25 & 14.54 & 1.44 & 1.09 & 0.280 & 0.260 & - & - & 1.37 & - & 4830 & [4], [3] \\ [2pt]
Mocs            & L5-6 & fall & 23.16 & 17.95 & 15.50 & 1.24 & 1.16 & 0.310 & 0.290 & 0.512 & 0.058 & 1.32 & - & 4680 & [5] \\ [2pt]
Knyahinya       & L/LL5 & fall & 20.80 & 19.17 & 15.00 & 1.33 & 1.28 & 0.406 & 0.270 & 0.746 & 0.090 & 1.16 & 0.21 & 4340 & [6], [7] \\ [2pt]
Ragland         & LL3.4 & find & 20.00 & 18.50 & 13.30 & 1.32 & 1.24 & 0.411 & 0.219 & 0.287 & 0.047 & 0.87 & 0.61 & 4240 & [8], [7] \\ [2pt]
Chelyabinsk     & LL5 & fall & 19.20 & 18.30 & 15.50 & 1.35 & 1.18 & 0.400 & 0.270 & 0.760 & 0.100 & 1.11 & - & 4330 & [9], [10]  \\ [2pt]
Kheneg Ljouâd   & LL5/6 & fall & - & - & - & - & - & - & - & - & - & - & - & 3820 & N/A \\ [2pt]
Eagle           & EL6 & fall & 23.19 & 20.19 & 14.43 & 0.54 & 1.80 & 0.220 & 0.220 & 0.616 & 0.075 & 1.38 & 0.32 & 4530 & [7] \\ [2pt]
Allende         & CV3 & fall & 23.16 & 15.94 & 15.47 & 1.84 & 1.82 & 0.404 & 0.150 & 0.349 & 0.038 & 1.40 & 0.29 & 4350 & [11] \\ [2pt]
Lancé           & CO3.5 & fall & 25.33 & 15.90 & 14.43 & 1.80 & 1.44 & 0.328 & 0.144 & 0.446 & 0.045 & 1.40 & 0.40 & 4610 & [12], [7] \\ [2pt]
Murchison       & CM2 & fall & 22.13 & 13.58 & 12.02 & 1.35 & 1.14 & 0.328 & 0.176 & 0.178 & 0.033 & 1.13 & 1.85 & 3860 & [7], [13] \\ [2pt]
Sariçiçek       & Howardite & fall & 14.40 & 23.75 & 9.94 & 5.34 & 3.77 & 0.756 & 0.460 & 0.255 & 0.025 & 0.05 & - & 4050 & [14] \\ [2pt]
Stannern        & Eucrite & fall & 13.93 & 23.23 & 4.21 & 7.61 & 6.49 & 0.220 & 0.400 & 0.430 & 0.061 & 0.0002& - & 4250 & [15], [16] \\ [2pt]
Bilanga         & Diogenite & fall & 12.20 & 24.80 & 18.00 & 0.50 & 0.44 & 0.601 & 0.348 & 0.009 & 0.011 & 0.002 & - & 3680 & [12], [17] \\ [2pt]
Tissint         & Shergottite & fall & 15.34 & 21.60 & 10.31 & 4.65 & 2.57 & 0.417 & 0.400 & 0.534 & 0.017 & 0.04 & - & 4830 & [18], [19] \\ [2pt]
NWA 11303       & Lunar & find & 2.76 & 20.68 & 2.61 & 12.01 & 15.66 & 0.050 & 0.050 & 0.250 & 0.020 & - & - & 4610 & [20]$^a$ \\ [2pt]
Norton County   & Aubrite & fall & 2.37 & 24.20 & 25.94 & 0.85 & 0.15 & 0.068 & 0.120 & 0.163 & 0.010 & 0.06 & - & 4000 & [21] \\ [2pt]
Dhofar 1575     & Ureilite & find & 14.76 & 18.47 & 21.25 & 0.77 & 0.28 & 0.522 & 0.310 & 0.063 & 0.025 & 0.13 & 2.80 & 3790 & [22]$^b$ \\ [2pt]
Mincy           & Mesosiderite & find & - & 22.19 & 11.02 & 3.78 & 3.54 & 0.523 & 0.410 & 0.163 & 0.007 & 0.47 & - & 4690 & [23] \\ [2pt]
Mount Joy       & Iron & find & 92.72 & 1.09 & - & - & - & 0.005 & - & - & - & 5.68 & - & 3900 & [24], [25] \\ [2pt]
\hline \\[-20pt]
\end{tabular}}
\end{center}
\tablefoot{The tested meteorites are described by their name, classification, recovery type, the bulk composition of the main species in wt\%, and the estimated plasma temperature. The final column includes the source of the reported bulk chemical composition values. Bulk composition references: [1] \citet{2015M&PS...50..864O}, [2] \citet{1961Dyakonova}, [3] \citet{2022GeCoA.318...19P}, [4] \citet{2011M&PS...46..652M}, [5] \citet{1995AMR.....8..153M}, [6] \citet{1989GeCoA..53.2747K}, [7] \citet{1990Metic..25..323J}, [8] \citet{1986Metic..21..217R}, [9] \citet{2013GeocI..51..522G}, [10] \citet{Kirillov2021}, [11] \citet{1987SmCES..27...49J}, [12] \citet{KUSUNO20132.0239}, [13] \citet{2018GeCoA.239...17B}, [14] \citet{2019M&PS...54..953U}, [15] \citet{1973E&PSL..18..433M}, [16] \citet{1998M&PSA..33..197K}, [17] \citet{2012M&PS...47...72M}, [18] \citet{2013LPI....44.2683H}, [19] \citet{Liu2016}, [20] \citet{2021M&PS...56..206K}, [21] \citet{1985Metic..20..571E}, [22] \citet{1970SSRv...10..483V}, [23] \citet{1977cami.coll..445S}, [24] \citet{1969GeCoA..33..859W}, [25] \citet{2007GeCoA..71..760W}. Meteorites for which no chemical composition reference was found are denoted as N/A.  \\ 
$^a$Mean composition from 24 typical feldspathic lunar meteorites in [20]; $^b$Mean composition from six ureilites in [22].}
\end{table*}

\begin{table*}
\small\begin{center}
\caption{Spectral lines measured and analyzed within this work.}
\label{T3}
\vspace{0.1cm}
\resizebox{.9\textwidth}{!}{
\begin{tabular}{lll}
\hline\\[-5pt]
\multicolumn{1}{l}{Multiplet}& %
\multicolumn{1}{l}{Wavelength [nm]}& %
\multicolumn{1}{l}{Notes / presence in meteorites} \\ 
\hline\\ [-6pt]
Mg I-3 & 382.93; 383.23; 383.83 & Strong in aubrite, diogenite, ureilite. Faint in eucrite, howardite, shergottite, lunar. \\ [3pt]
CN (UV band) & 380 - 389; peak at 388.33 & Strong in carbonaceous chondrites, volatile-rich bodies, ureilite. \\ [3pt]
Si I-3 & 390.55 & Strong in HEDs, aubrite, lunar, shergottite. Faint in carbonaceous chondrites, ureilite. \\ [3pt]
Mn I-2 & 403.08; 403.32; 403.46 & Strong in enstatite chondrite, shergottite. Faint in carbonaceous chondrites, ureilite. \\ [3pt]
Cr I-1 & 425.45 & Strongest Cr I line, an alternative to Cr I-7 for low-resolution data. \\ [3pt]
Mn I-16 & 478.34; 482.35 & Fainter lines alternative to Mn I-2 for low-resolution data. \\ [3pt]
H I-1 & 486.13 & Strong in carbonaceous and volatile-rich bodies. \\ [3pt]
Mg I-2 & 516.75; 517.21; 518.30 & Strong in aubrite, diogenite, ureilite. Faint in eucrite, howardite, shergottite, lunar. \\ [3pt]
Cr I-7 & 520.45; 520.60; 520.84 & Strong in HEDs, LL chondrites, mesosiderite. \\ [3pt]
\multicolumn{1}{l}{\multirow{3}{*}{Fe I-15}} & 526.97; 532.89; 537.10; & \multirow{3}{*}{Strong in chondrites, mesosiderites, iron. Faint in diogenite, lunar, aubrite.} \\ 
\multicolumn{1}{l}{} & 539.71; 540.53; 542.97; &  \\ 
\multicolumn{1}{l}{} & 543.45; 544.69; 545.56 &  \\ [3pt]
Ni I-59 & 547.60 & Not detected in achondrites. \\ [3pt]
Na I-6 & 568.29; 568.85 & Strong in some LL and carbonaceous chondrites, faint in diogenite, ureilite, mesosiderite, iron. \\ [3pt]
Na I-1 & 588.97; 589.56 & Saturated in most laboratory meteorite data. \\ [3pt]
H I-1 & 656.28 & Strong in carbonaceous and volatile-rich bodies. \\ [3pt]
Li I-1 & 670.81 & Faint or missing in chondrites.   Strong in eucrite, ureilite. \\ [3pt]
K I-1 & 766.53; 769.94 & Strong in some LL and carbonaceous chondrites, faint in HEDs, ureilite, mesosiderite, iron. \\ [1pt]
\hline \\[-20pt]
\end{tabular}}
\end{center}
\tablefoot{Atomic multiplet numbers, approximate wavelength positions, and notes related to the presence of these spectral features in various meteorite groups are given.}
\end{table*}

\section{Spectral features of ablated meteorite samples} \label{results}

To estimate the temperatures of the emitting plasma generated during the meteorite ablation, we have applied a simple spectrum model assuming local thermal equilibrium and optically thin radiating plasma. The model was used to fit most Fe I lines in the 406 -- 546 nm range for each meteorite. As a precaution, the strongest Fe lines were omitted from the fit. The resulting plasma temperatures determined using the Boltzmann plot method ranged between 3680 and 4830 K (Table \ref{T1}). The average uncertainty of the estimated temperatures is around 200 K. This range is consistent with the temperatures of most meteors. The typical temperature of the main (sometimes called low-temperature) component in meteors is usually estimated at $\approx$ 4500 K \citep{1994P&SS...42..145B}.

The reasons for the temperature differences between individual meteorites, considering that the plasma conditions during the experiments were the same, are not yet fully resolved. Variations in bulk composition and sample sizes may affect the resulting temperatures. We however note that significant temperature differences were estimated for a few meteorites with similar composition (Table \ref{T1}). The temperature estimate is also sensitive to the fitted wavelength range and selection of measured Fe lines, as not all may fully satisfy the thermal equilibrium assumption. Furthermore, although the radiating plasma appears close to optically thin, self-absorption may have affected the intensities of some fitted lines. A detailed analysis of the temperatures and relative element abundances in the meteorite plasma using a more comprehensive radiative-transfer model is planned for a separate study.

In general, an accurate evaluation of the composition of the radiating meteoric plasma requires fitting emission spectra based on a relevant radiative transfer model accounting for emission and absorption within the often optically thick plasma cloud. Such an approach has been previously applied to analyze several fireball spectra \citep{1993A&A...279..627B, 2003M&PS...38.1283T, 2007AdSpR..39..491J, 2018A&A...610A..73F, 2022MNRAS.514.5266K}, though some limitations in the ability to reproduce the expected element abundances have been noted, likely due to incomplete evaporation of some refractory elements.

Such analysis can be time-consuming, especially when interpreting large datasets of meteor spectra. Moreover, it can pose difficulties in determining accurate plasma condition parameters when using lower-resolution spectral data. Here we focus on identifying apparent spectral features that can be resolved individually in meteor spectral data. Such features include multiplet or line intensity ratios which exhibit notable variations between different meteorite composition types, making them suitable diagnostic tools for better evaluation of the meteoroid composition. Furthermore, as noted earlier, the results of our radiative transfer modeling of several meteorite Echelle spectra suggest that the radiating plasma is close to optically thin, meaning that self-absorption plays a less significant role than in bright meteors. This potentially allows for a better correlation between measured line intensities and corresponding relative atomic abundances in our dataset, as demonstrated in the next section. The list of spectral features further analyzed in this work is given in Table \ref{T3}.

\subsection{Overall characteristics}

\begin{figure*}
    \centering
\begin{subfigure}{0.48\textwidth}
    \includegraphics[width=\textwidth]{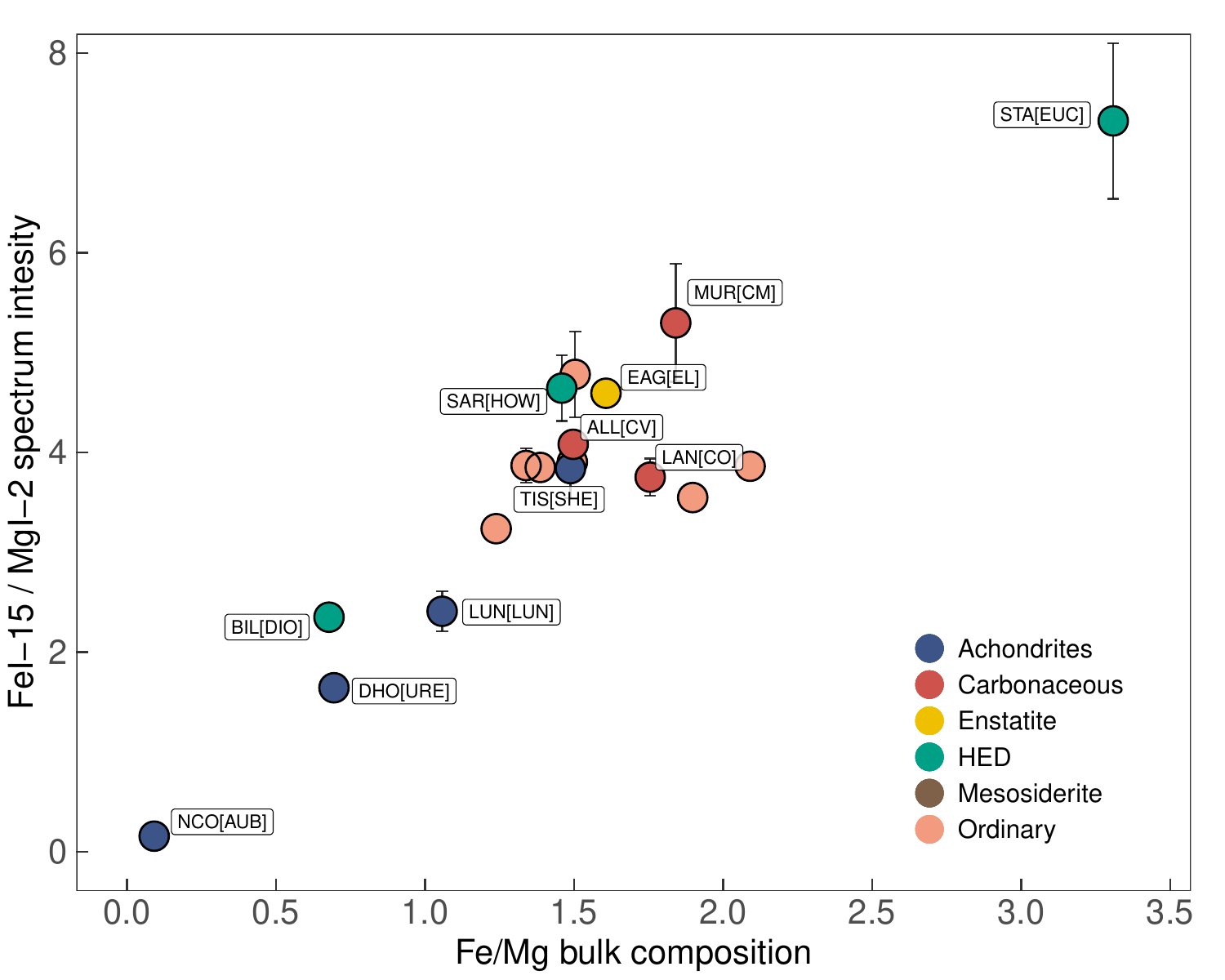}
    \caption{}
    \label{bulk_a}
\end{subfigure}
 \begin{subfigure}{0.48\textwidth}
    \includegraphics[width=\textwidth]{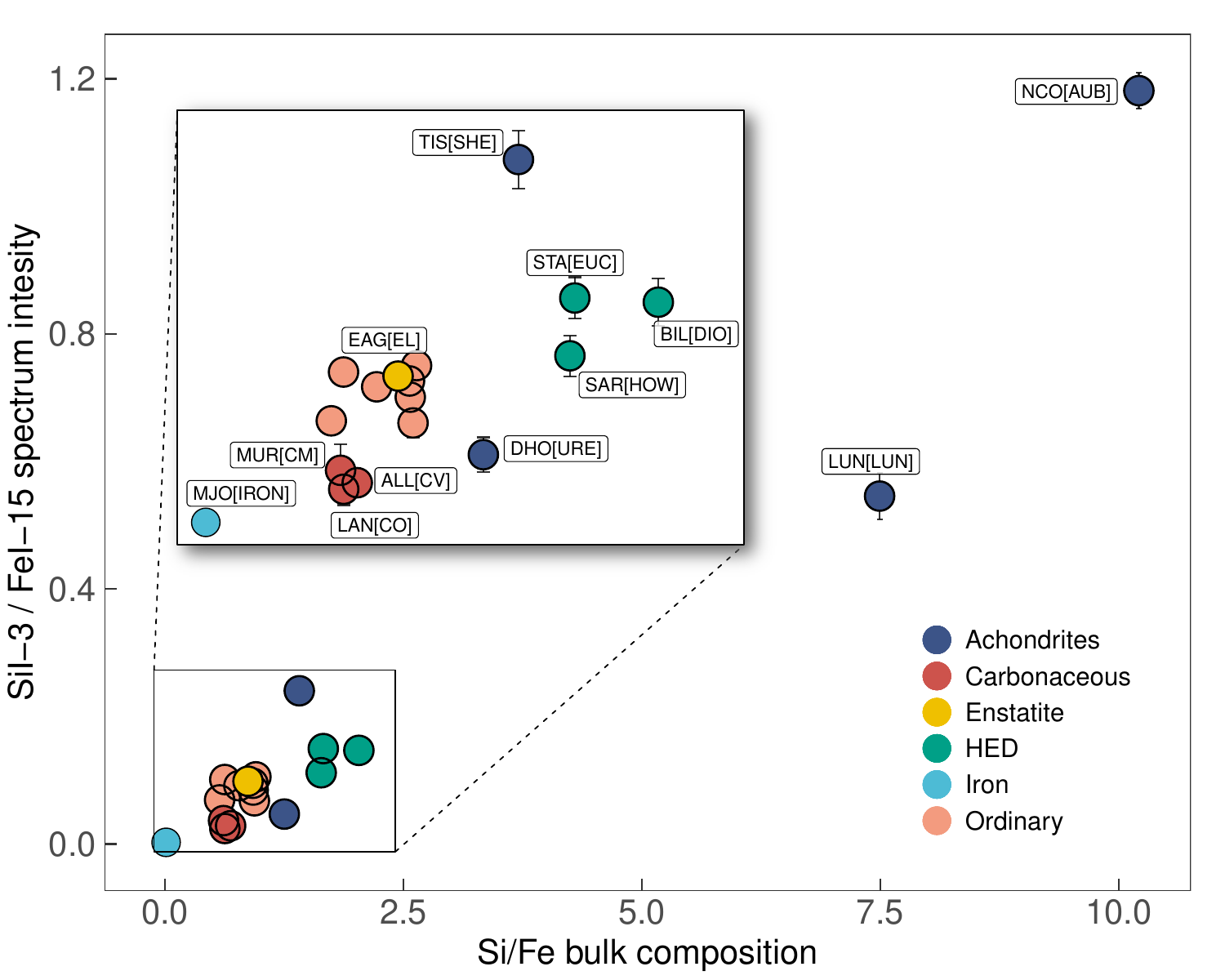}
    \caption{}
    \label{bulk_b}
\end{subfigure}
 \begin{subfigure}{0.48\textwidth}
    \includegraphics[width=\textwidth]{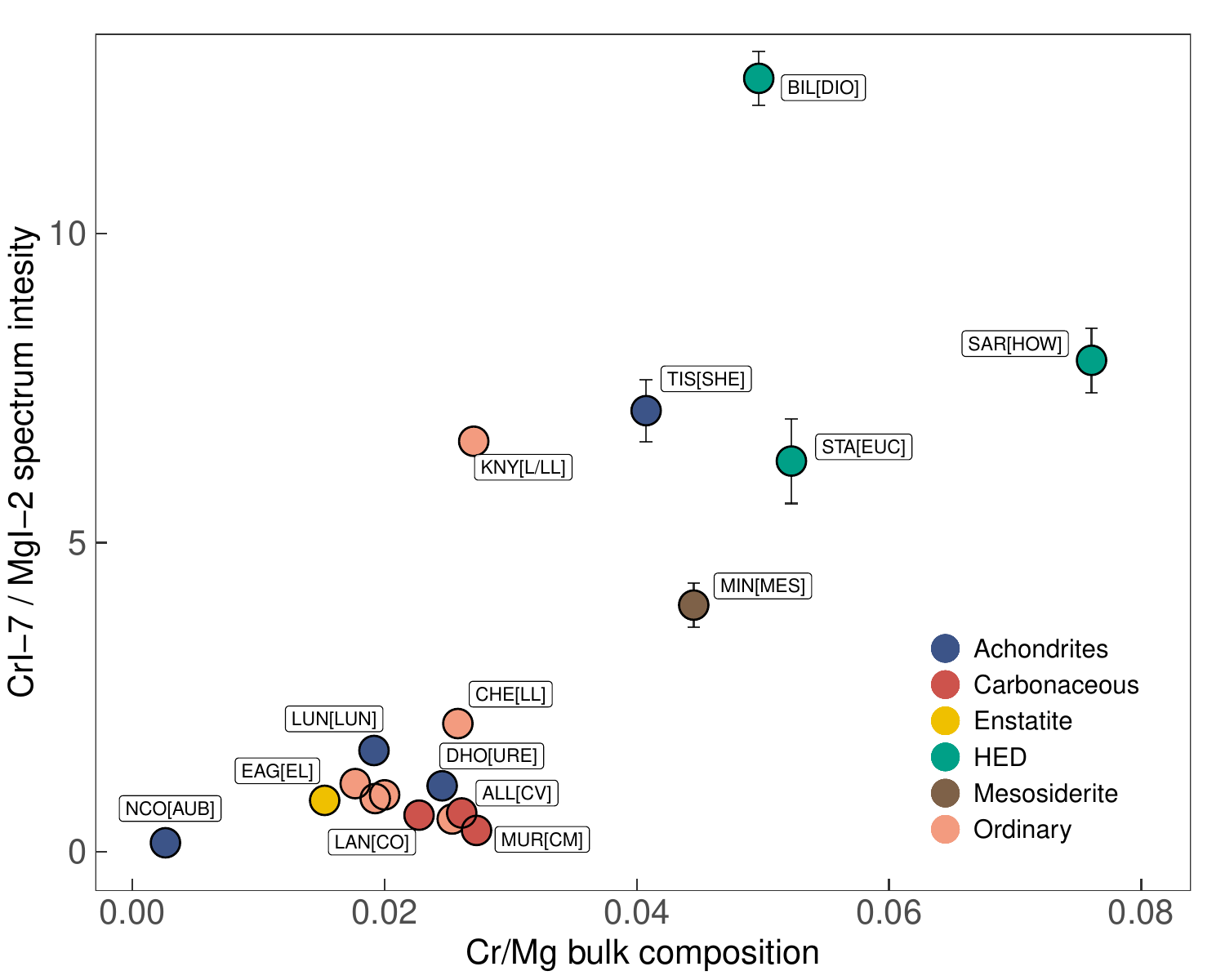}
    \caption{}
    \label{bulk_c}
\end{subfigure}
 \begin{subfigure}{0.48\textwidth}
    \includegraphics[width=\textwidth]{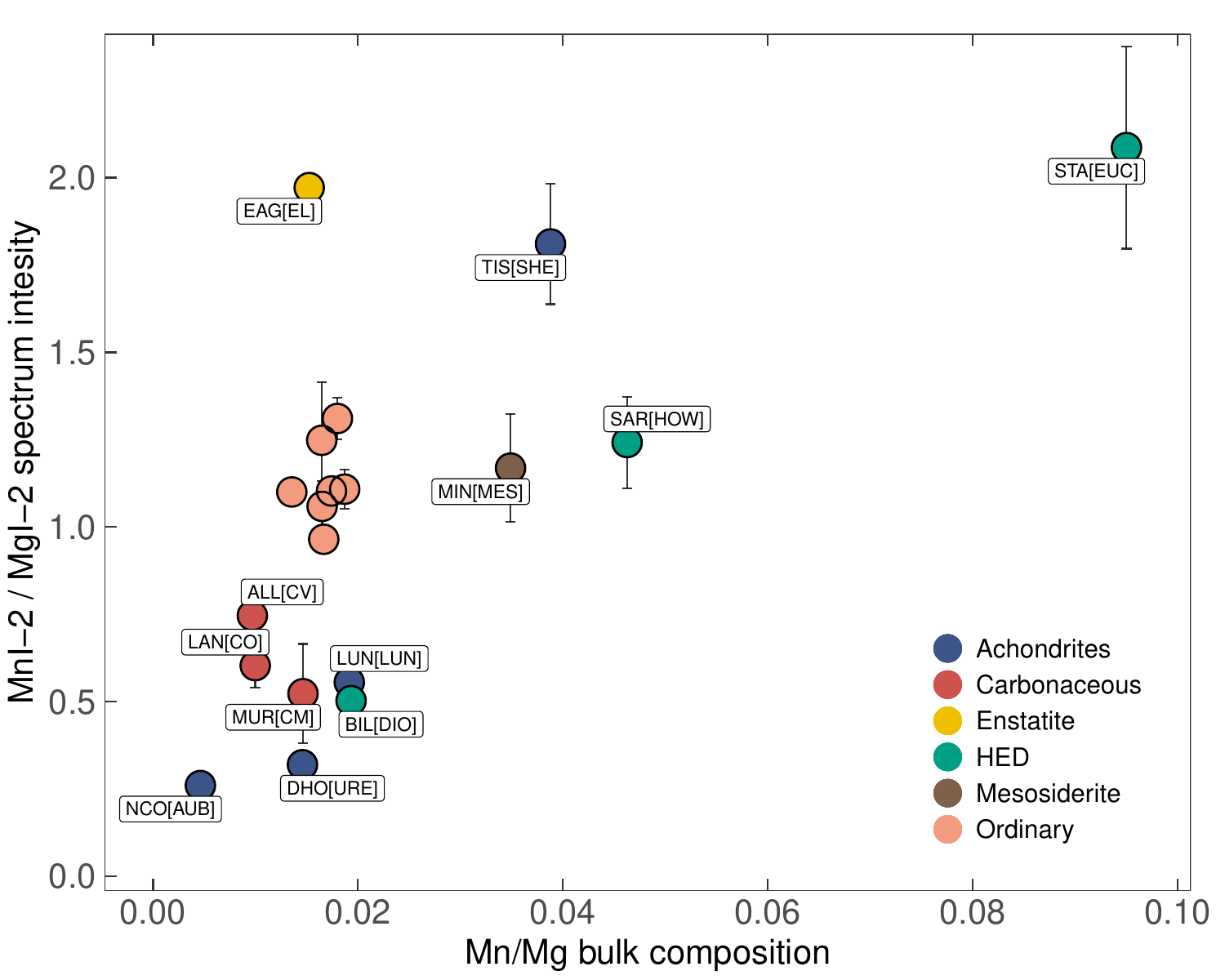}
    \caption{}
    \label{bulk_d}
\end{subfigure}
\caption{Correlation between the assumed original meteorite bulk chemical composition (Table \ref{T1}) and the measured intensities of corresponding emission lines in spectra of ablated meteorites. Meteorite labels containing three letters from the meteorite's name and classification are given for all meteorites except most ordinary chondrites. In all figures, the HED achondrites are plotted separately from the rest of the achondrites. In Fig. \ref{bulk_c}, meteorite Ragland (LL chondrite) is outside the displayed limits with Cr I-7 / Mg I-2 intensity $\approx$ 25.}
 \label{bulk}
\end{figure*}

Since we aim to identify spectral line intensity ratios that can help constrain the specific meteorite/meteoroid composition type, it is important to validate that these features trace the original composition of the sample. Most of the tested meteorites have been previously analyzed by various techniques in the laboratory, and published references for their bulk chemical composition are available. This allows us to confront the measured spectral differences between meteorites with the variations of their original chemical composition. Examples of the correlation between these two datasets are displayed in Fig. \ref{bulk}. The presented emission line multiplets are among the most studied in meteor spectra analyses, as they are easily detectable by most meteor spectrographs operating in the visible range. The diagnostic spectral lines described within this work were selected based on their ability to reflect specific meteorite types and the accessibility to be measured in meteor spectral data.

The results in Fig. \ref{bulk} demonstrate that the meteorite bulk chemical composition variations are clearly reflected in the intensities of the analyzed spectral lines. Meteorites with a distinct composition in the given metric (element bulk content ratio) can be easily recognized based on their spectra. However, differentiating meteorites with relatively similar chemical composition in the given metric can be difficult based on a single spectral feature. An example of this can be seen when attempting to distinguish ordinary, carbonaceous, and enstatite chondrites based on their Fe/Mg or Cr/Mg spectrum intensity (Fig. \ref{bulk_a} and \ref{bulk_c}). This mixing of various meteorite classes in specific criteria has several causes. 

Firstly, the meteorite classification is based on various mineralogical, chemical, and petrological aspects. Some meteorite classes can therefore be chemically heterogeneous and different meteorite types can overlap for specific element abundance ratios. When interpreting the results in Fig. \ref{bulk}, it should be remembered that while the presented bulk composition data are based on reliable references for the same meteorite cases, different samples were used for the bulk chemical composition analyses and the ablation analyses with collected emission spectra. Minor chemical heterogeneities can be expected between individual samples even for the same meteorite case. Finally, we note that the composition of the radiating meteor plasma will not fully reflect the chemical composition of the original meteorite/meteoroid, due to the presence of refractory elements, incomplete evaporation of certain species at given plasma temperatures, and the lack of strong atomic transitions for some of the studied elements in the recorded wavelength range \citep{1993A&A...279..627B}.

From the point of view of spectral measurements, several factors can affect the ability of a spectral multiplet to reflect the corresponding atom abundance. The accuracy of the line intensity measurements in the presented Echelle spectra is very high (the ratio uncertainties in Fig. \ref{bulk} are mostly within the sizes of the displayed data points). However, the measurement accuracy of some spectral multiplets will be lower in meteor spectra analyses often based on lower-resolution data affected by line blending and saturation. Furthermore, besides the atom concentrations, the conditions of the radiating plasma affect the observed spectral signatures and can vary more significantly in meteors. A more straightforward correlation between the original chemical composition and the relevant measured line intensities will be observed at conditions satisfying the local thermodynamic equilibrium and without significant self-absorption within the plasma cloud. Different line multiplets will be affected by the specific plasma conditions to a different degree. These effects can be quantitatively characterized by constraining the plasma conditions using a radiative transfer model fitted to the spectrum in a wider wavelength range.

\subsection{Diagnostic spectral features}

\subsubsection{Spectral features from previous meteor analyses} \label{sectionPrevious}

The most commonly measured spectral features in meteors are emission lines of Fe I, Mg I, Na I, and Ca II. The relative intensities of the most prominent emission line multiplets of these species have often been used to characterize meteor spectra and provide a broad interpretation of their composition \citep[][and others]{1998SSRv...84..327C,2005Icar..174...15B, 2019A&A...629A..71M}. Besides O I and N I lines, which mainly originate from the heated atmosphere, these species comprise the strongest lines observable in most meteors. Some compositional variations can be revealed already by studying these strongest features. Iron meteoroids are easily recognized, as Fe I lines dominate their spectra and other species are often absent or very weak. The variations of the relative intensity of the strongest Na I multiplet (Na I - 1) have been used to reveal the effects of space weathering causing the loss of volatiles in meteoroids \citep{2005Icar..174...15B, 2019A&A...629A..71M, 2020P&SS..19405040A}. The interpretation of the intensities of some emission lines (particularly low-excitation lines such as Na I-1) needs to be considered carefully. The differences in the physical conditions of the meteor ablation, particularly the plasma temperature related to the meteor speed, may affect the strength of these lines more significantly than the embedded compositional differences between meteoroids \citep{2020Icar..34713817M}. In the abovementioned studies, the measured relative intensities of the main multiplets of Fe I-15, Mg I-2, and Na I-1 have been compared with the expected values corresponding to CI chondritic composition to evaluate the compositional variations of the studied meteoroids. 

These previous efforts have distinguished several meteor spectra classes, which are assumed to correspond to rough compositional characteristics of meteoroids. The chondritic-like meteoroids (also classified as normal-type spectra) comprise the majority of previously analyzed meteors \citep{2019A&A...621A..68V, 2019A&A...629A..71M} and likely include both cometary meteoroids and asteroidal meteoroids with a composition similar to carbonaceous and ordinary chondrites. Based on the results presented in this work, using these spectral criteria (Fe I-15, Mg I-2, and Na I-1), meteoroids similar to enstatite chondrites, some mesosiderites, and some types of achondrites could also be classified into this group. The meteoroids with depleted volatile content (classified as Na-poor and Na-free spectra) mainly reflect the thermal alteration and meteoroid weathering degree, rather than their original chemical composition characteristics. Meteoroids with potentially increased volatile content may be present among the proposed Na-enhanced and Na-rich spectral classes \citep{2005Icar..174...15B}, but we suspect that many of the meteoroids previously identified within this spectral group were affected by the low speed of these meteors, rather than their composition \citep{2020Icar..34713817M}. This classification also identified an iron spectral type, which directly corresponds to iron meteoroids. Additionally, in \citet{2019A&A...629A..71M} we have proposed Fe-rich spectral type within this classification, positioned between the normal and iron classes and potentially corresponding to chondrites with enhanced iron content and stony-iron bodies.

The above summary aimed to demonstrate that while some aspects of meteoroid composition can be recognized using the commonly studied major spectral features in meteors, further spectral analysis of other species is required to discern the true diversity of meteoroid materials, partly consisting of the meteorite classes found on Earth. In the following text, we describe additional spectral features that can help constrain specific meteoroid composition types, as identified in our analysis of the higher-resolution spectra of ablated meteorite samples. The identified spectral species are discussed in sections grouped based on their observed relative differences and approximate wavelength range, not reflecting any mineralogical links.

\subsubsection{Cr I, Mg I, Fe I}

Simple distinction based on the measured intensity ratios of Fe I / Mg I can already help differentiate between low-Fe materials (e.g., aubrites or lunar meteorites), high-Fe materials (irons, stony-irons), and bodies with comparable contents of Fe and Mg (ordinary, carbonaceous and enstatite chondrites). However, as evident in Fig. \ref{bulk_a}, compositionally distinct meteorites may exhibit similar Fe I / Mg I intensity ratios. For example, if we only consider this spectral measure, the Martian meteorite or the howardite would appear comparable to ordinary chondrites. Furthermore, carbonaceous and ordinary chondrites also cannot be distinguished this way. The interpretation of more distinct groups can also be problematic, as it may not be clear if we are looking at a Mg-poor (such as the eucrite) or Fe-enhanced material. 

\begin{figure}
\centerline{\includegraphics[width=.98\columnwidth,angle=0]{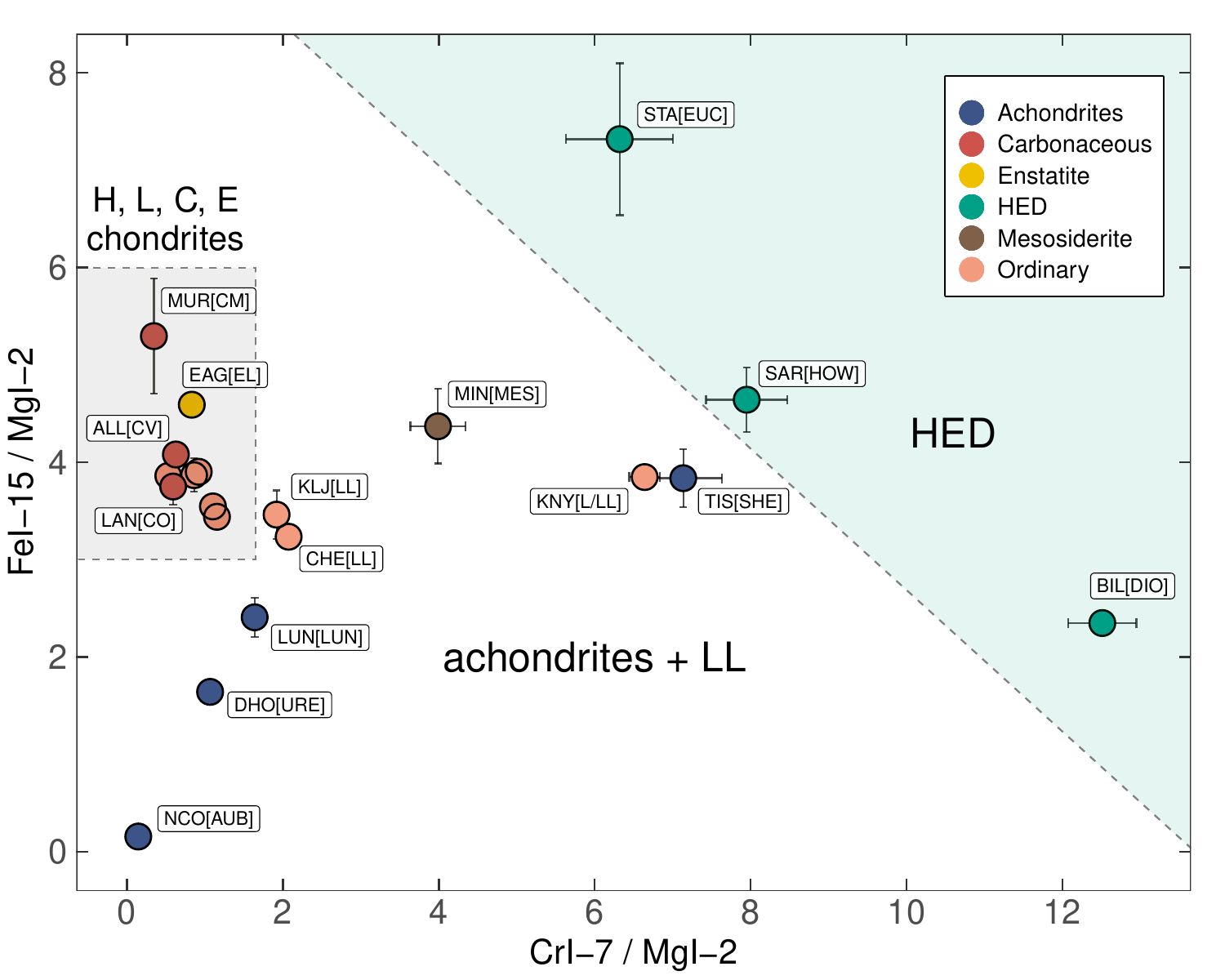}}
\centerline{\includegraphics[width=\columnwidth,angle=0]{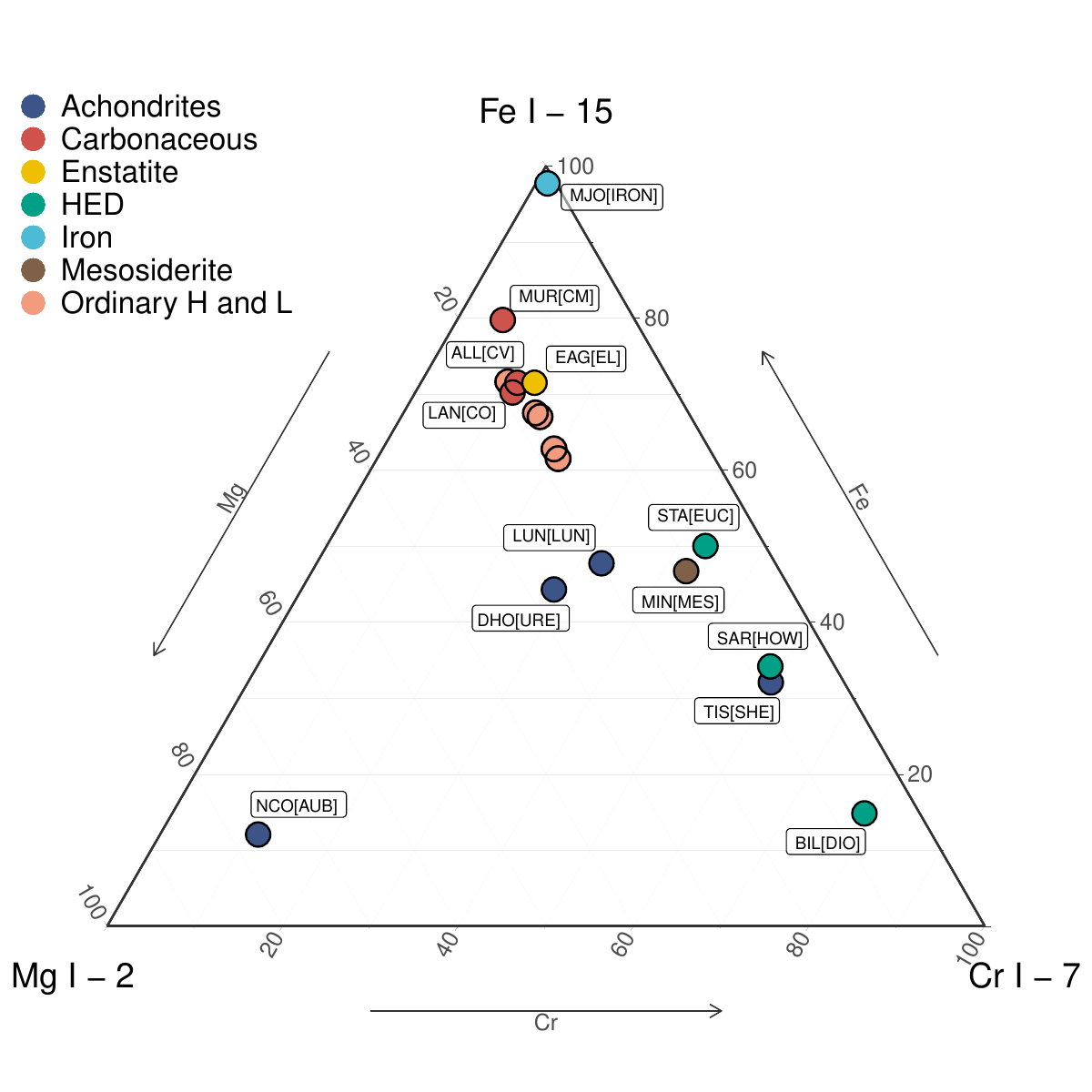}}
\caption[f1]{Distinction of the ablated meteorite samples simulating asteroidal meteor analogs based on the measured relative intensity of Cr I-7, Mg I-2, and Fe I-15 spectral lines. Upper panel: Dependency between Cr I and Fe I line intensity relative to Mg I for all tested meteorites. Regions characteristic of HEDs and chondrites (except for the LL group) are highlighted. The Ragland (weathered LL3.4; Cr/Mg $\approx$ 25) and Mount Joy (iron; Fe/Mg $\approx$ 100) meteorites are outside the displayed plot limits. Lower panel: ternary diagram displaying relative intensities of Cr I, Mg I, and Fe I for all samples except the LL group.}
\label{CrMgFe}
\end{figure}

We have found that additional measurement of the nearby Cr I lines can help constrain specific meteorite composition types. Chromium is a relatively minor component in meteorites (Table \ref{T1}) but is represented by several strong lines in meteor spectra \citep{1994A&AS..103...83B}. In our analysis of the spectra of ablated meteorites, we have particularly focused on the Cr I-7 multiplet consisting of three lines near 520.4 -- 520.9 nm, often seen blended into one strong line in lower-resolution spectra. Fig. \ref{CrMgFe} shows results from our measurements of Cr I intensity relative to the nearby Fe I and Mg I multiplets. Based on this metric, the basaltic HED (Howardite-Eucrite-Deiogenite) meteorites can be distinguished by increased Cr I / Mg I intensities, reflecting the relatively high Cr and low Mg composition of these meteorites. Other distinct achondrites can also be identified, such as the Fe-poor and Mg-rich aubrite. The ternary diagram in Fig. \ref{CrMgFe} shows that comparable relative intensities of the Cr I, Mg I, and Fe I were found in the ureilite and lunar meteorites. 

Furthermore, all of the tested chondrites (carbonaceous, ordinary, and enstatite) form a relatively compact group with low  Cr I / Mg I intensities and similar Fe I / Mg I (Fig. \ref{CrMgFe}). The exception to this trend is presented by LL chondrites, which were found to be more heterogeneous than other meteorite classes, with generally higher Cr I / Mg I intensities than other chondrites. As will be apparent in the next sections, this is a recurring trend also for other diagnostic spectral features. The LL group appears spectrally more heterogeneous than other chondrites, making its characterization based on a specific spectral signature difficult. Further discussion about this group is given in Section \ref{Ordinary_discussion}.

\subsubsection{Si I, Mn I}

Silicon is one of the main components in the chemical composition of most meteorites except for irons. Yet, analyzing Si content from visible meteor spectra is rather difficult. In previous analyses, only the two lines of ionized Si II at 534.7 and 537.1 nm (Si II-2) were typically reported. These lines are not detected in slower meteors as they are assumed to originate from the high-temperature spectral component. Only one line of neutral Si I has been previously detected in meteor spectra -- the Si I-3 line near 390.55 nm. However, this line has been rarely analyzed in meteor surveys, due to the difficulty of its detection in lower-resolution data where it is blended with nearby strong Fe I and Ca II lines.

Nearby wavelength region also includes the strongest spectral multiplet of manganese -- the Mn I-2 line triplet (403.08, 403.31, 403.45 nm). Manganese bulk content in meteorites is similar to chromium (Table \ref{T1}), ranging from 0.19 to 0.46 wt$\%$ in the investigated samples. The Mn I-2 lines were previously reported in higher-resolution meteor spectra, but have not been commonly analyzed. A resolving power of R > 500 is needed to reliably recognize the contribution of Mn I-2 lines from the strong Fe I-43 peak near 404.58 nm.

\begin{figure}
\centerline{\includegraphics[width=.98\columnwidth,angle=0]{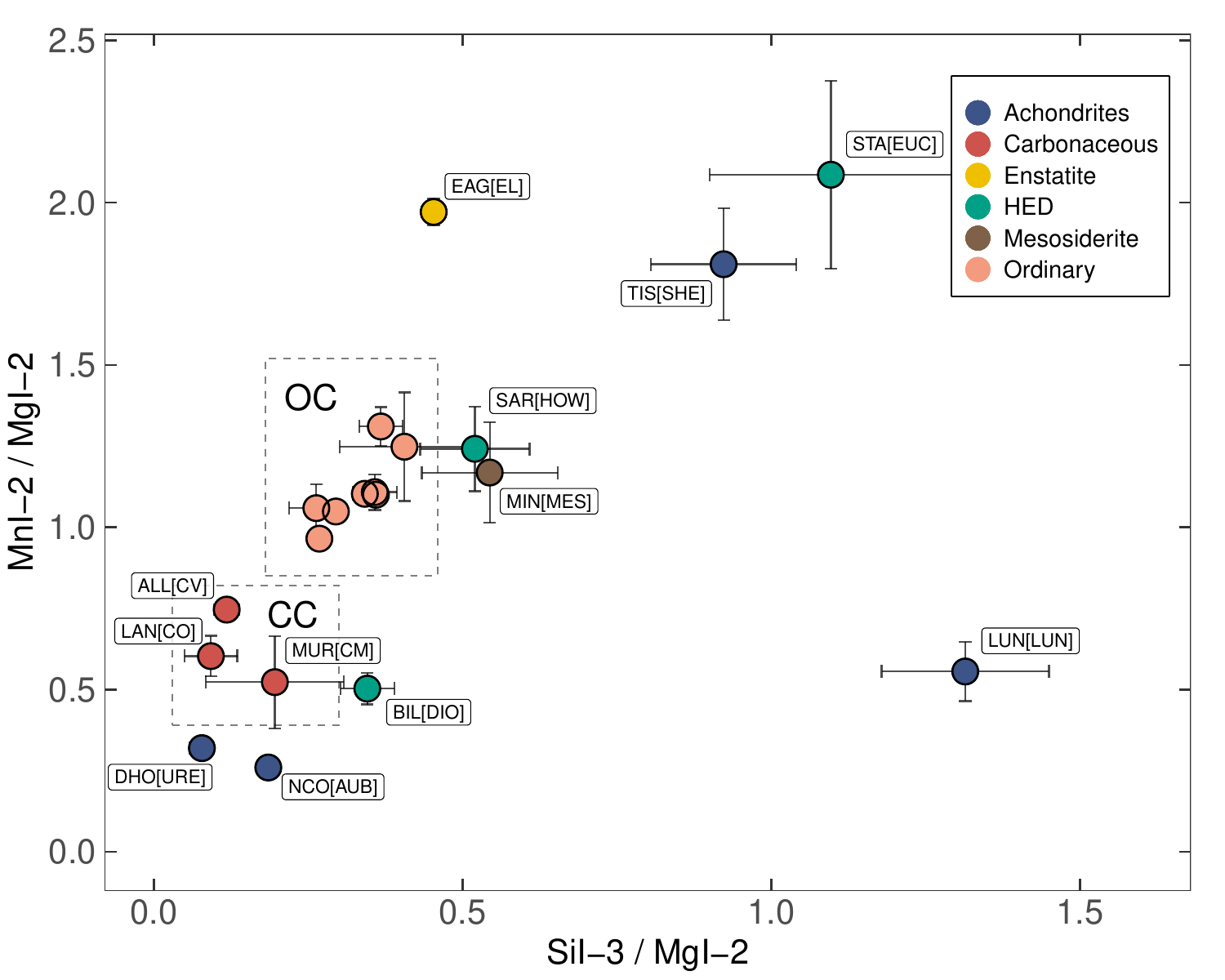}}
\centerline{\includegraphics[width=\columnwidth,angle=0]{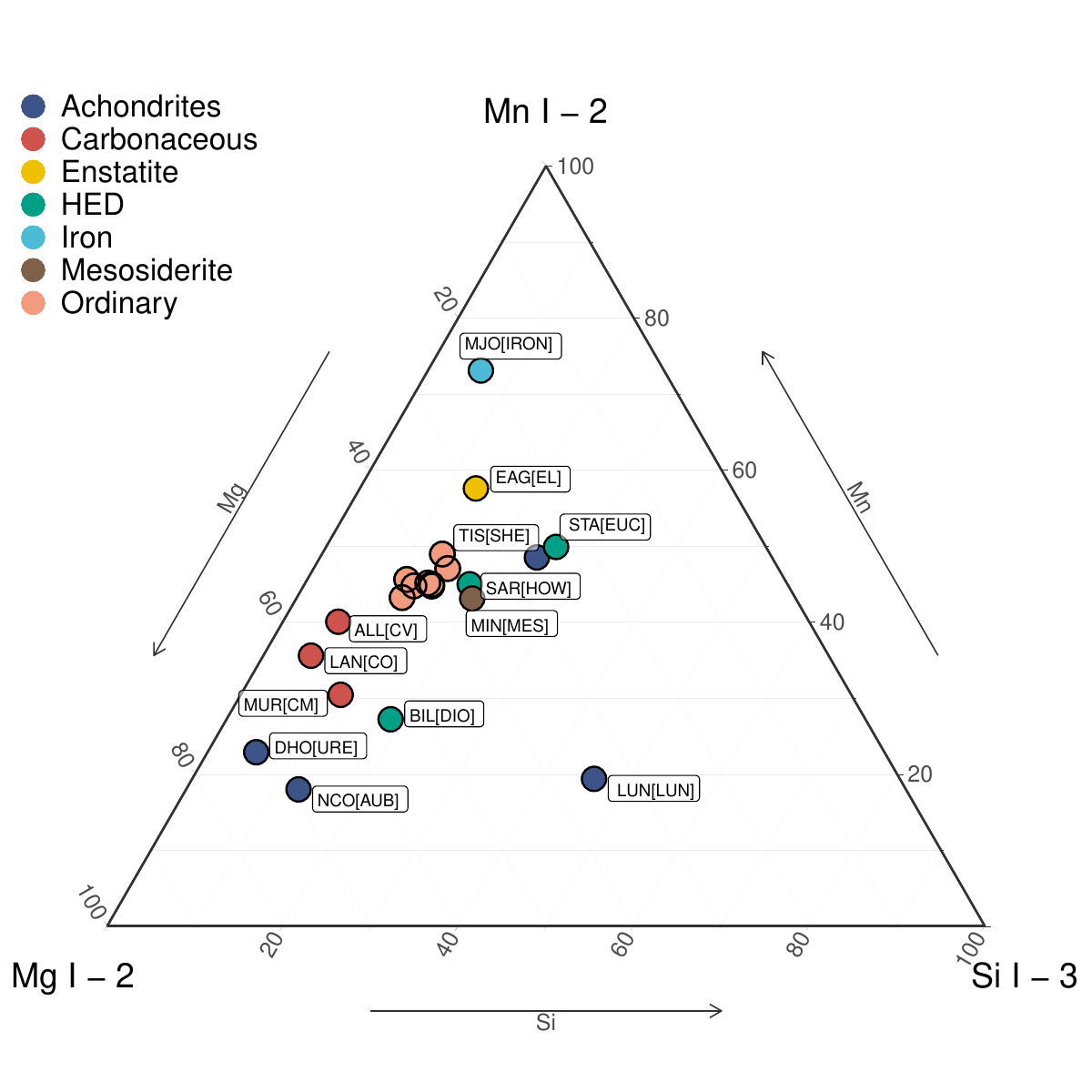}}
\caption[f1]{Measured relative intensities of Si I-3, Mg I-2, and Mn I-2 in spectra of ablated meteorite samples. Upper panel: Dependency between Si I and Mn I line intensity relative to Mg I for all tested meteorites. Approximate sections characteristic of ordinary (OC) and carbonaceous (CC) chondrites are highlighted. The Mount Joy (iron; Mn/Mg $\approx$ 3.5) meteorite is outside the displayed plot limits. Lower panel: Ternary diagram displaying relative intensities of Si I, Mg I, and Mn I for all samples.}
\label{SiMg_MnMg}
\end{figure} 

\begin{figure}
\centerline{\includegraphics[width=.98\columnwidth,angle=0]{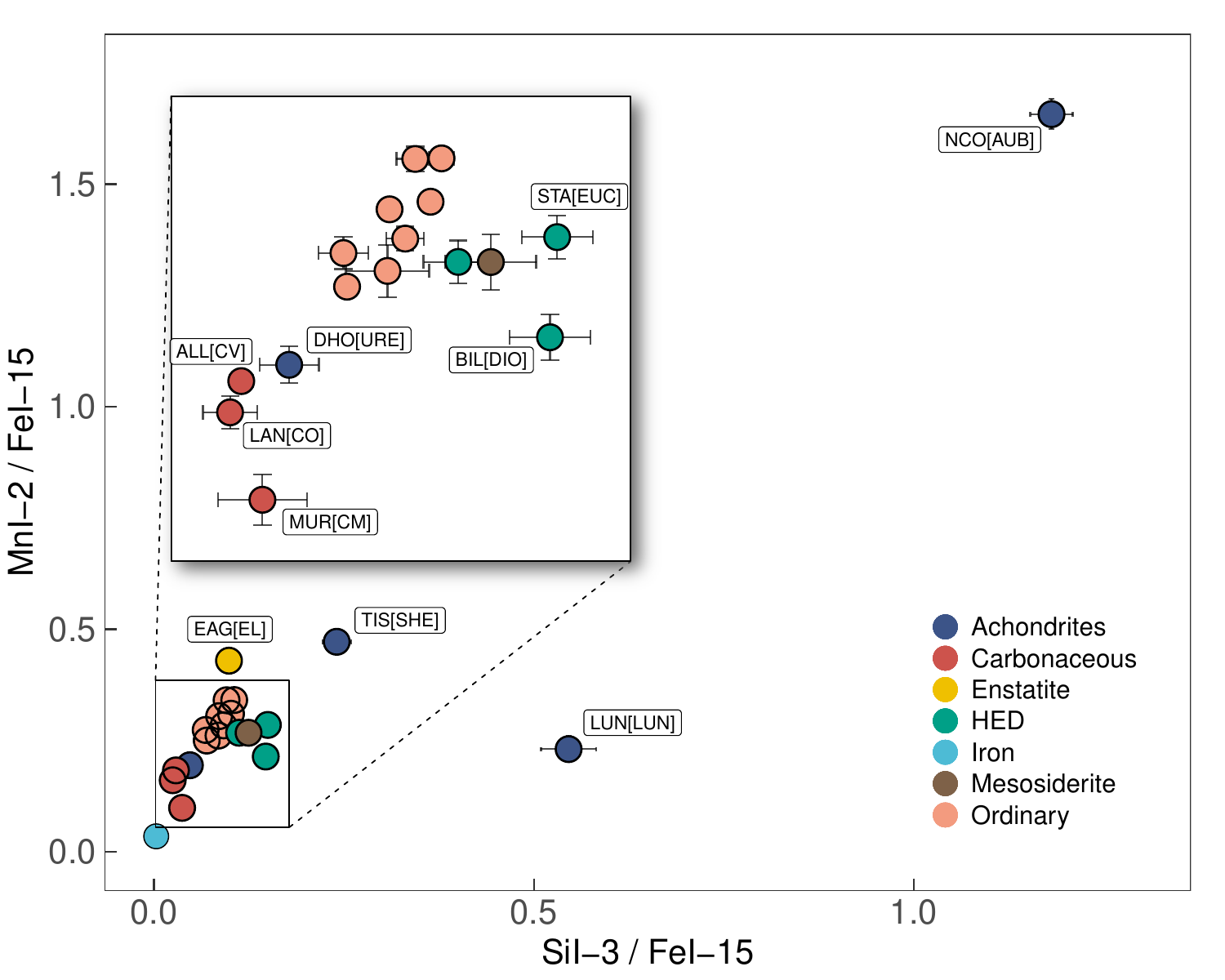}}
\caption[f1]{Correlation between the measured Si I-3 and Mn I-2 line intensities relative to Fe I-15 in spectra of meteorite samples.}
\label{SiFe_MnFe}
\end{figure}

Our results from the meteorite ablation analysis suggest that both Si I-3 and Mn I-2 are suitable diagnostic tools for constraining specific meteoroid composition types. The results displayed in Fig. \ref{SiMg_MnMg} imply that ordinary chondrites can be constrained by similar Mn I / Mg I and Si I / Mg I intensity ratios, while carbonaceous chondrites exhibit fainter emission for both Si I and Mn I. The plot in Fig. \ref{SiMg_MnMg} may be useful for identifying ordinary chondrites, which form a more heterogeneous group (particularly LL chondrites) in most of the other presented spectral features. The Si I and Mn I intensities measured relative to Mg I or Fe I (Fig. \ref{SiFe_MnFe}) are among the few spectral features in which ordinary, carbonaceous, and enstatite chondrites are clearly distinguished, besides the intensities of species tracing the presence of hydrated minerals and organic compounds (H I, CN, see Section \ref{HCNtext}).

Different types of achondrites can also be recognized using this spectral metric. The faintest Si I and Mn I emission was observed from the ureilite. The distinct position of the aubrite is caused by its strong Mg I and faint Fe I lines (Figs. \ref{SiMg_MnMg} and \ref{SiFe_MnFe}). The lunar meteorite can also be easily identified, as it exhibits the largest Si I / Mg I intensity ratio and among the lowest intensities of Mn I lines. Both these spectral features correspond with the bulk chemical composition of the lunar meteorite NWA 11303 (Table \ref{T1}). Strong emission of both Mn I and Si I, while weaker Mg I lines were observed from the Martian shergottite Tissint. Relatively high manganese contents appear to be characteristic of the Martian surface, as evidenced by in-situ measurements and SNC (shergottite-nakhlite-chassignite) meteorite studies \citep{McSweenJrTreiman+1998+953+1006, 2014GeoRL..41.5755L}. Based on Fig. \ref{SiMg_MnMg}, the HED achondrites may appear to have heterogeneous Mn I and Si I intensities. This is instead mainly caused by their considerably varying Mg content (Table \ref{T1}), which directly correlates with the intensity of Mg I-2 in their spectra. When compared relative to the Fe I emission, the measured Mn I and Si I intensities are relatively similar among the studied HEDs, with Si/Fe intensities enhanced over the values typical for ordinary chondrites (Fig. \ref{SiFe_MnFe}). 

\subsubsection{Li I}

While lithium is a minor component of meteorites, its isotopes are sensitive indicators of aqueous alteration processes and can be useful for recognizing terrestrial processes owing to their redistribution in the presence of aqueous fluids \citep{2004ApJ...612..588S}. Studies based on individual meteorite samples estimated bulk Li concentrations mainly between 1.4 and 2.0 ppm in ordinary chondrites \citep{1968E&PSL...5...17B, 2022GeCoA.318...19P}, between 1.4 and 2.4 in most carbonaceous chondrites \citep{2022GeCoA.318...19P}, between 1.3 ppm (Chassigny) and 5.6 ppm (Shergotty) in the Martian meteorites \citep{1982LPI....13..186D}, or around 0.33 ppm in the aubrite Norton County \citep{1968E&PSL...5...17B}. Higher concentrations of Li were measured in the basaltic HED meteorites. For HED samples analyzed in this work, they range from 2.9 ppm in the diogenite Bilanga to 10.9 ppm in the eucrite Stannern \citep{2014GeCoA.125..131M}. Increased Li concentrations ranging from 5 to 49 ppm were measured in lunar samples \citep{2006E&PSL.245....6S, 2021M&PS...56.1829F}.

\begin{figure}
\centerline{\includegraphics[width=.98\columnwidth,angle=0]{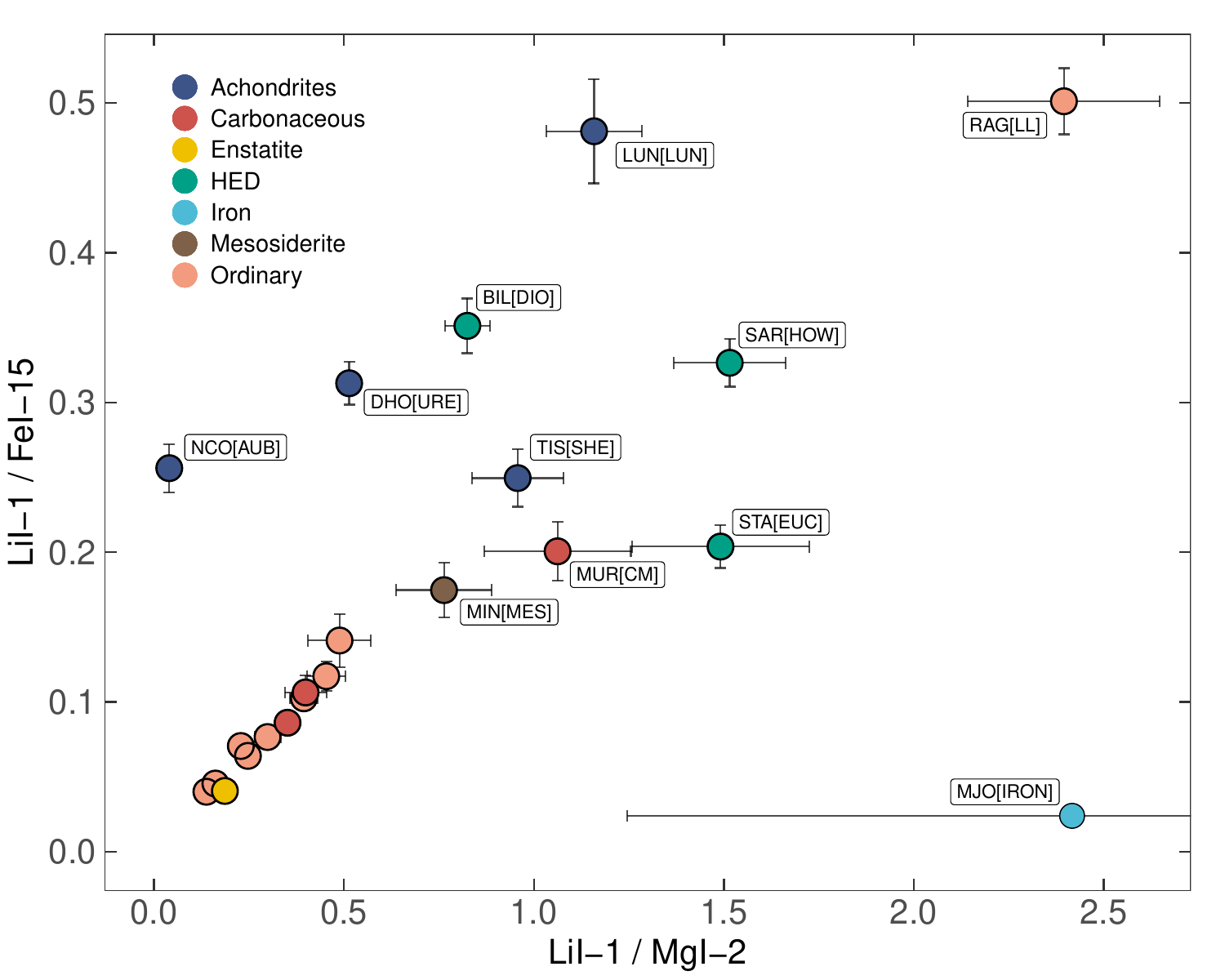}}
\caption[f1]{Li I - 1 line intensities measured relative to Fe I - 15 and Mg I - 2 in spectra of meteorite samples.}
\label{LiMg_LiFe}
\end{figure} 

The estimated Li concentration variations are well-reflected in the measured Li I-1 (670.79 nm) line intensities within our dataset (Fig. \ref{LiMg_LiFe}). The chondrites form a compact group with faint Li I emission, except for the CM chondrite Murchison showing a moderate increase of Li, and the weathered LL chondrite Ragland showing the strongest Li I line. Both these samples are affected by alteration. The aubrite sample Norton County exhibited moderate Li I emission, but its position in Fig. \ref{LiMg_LiFe} is again affected by its distinct Fe-poor and Mg-rich composition. Other achondrites are quite widely distributed using this metric, with overall stronger Li emission detected in the HED, Martian, and lunar meteorite samples, in correlation with their estimated original chemical composition. 

Our results imply that the measurement of the Li I-1 intensity can be used to distinguish chondritic-like interplanetary materials from more differentiated meteoroids, and potentially even recognize different types of achondrites. Our measurements of Li emission from ablated meteorites are in agreement with the recent study of two fireball spectra linked with recovered meteorites. \citet{2024A&A...686A..67S} compared the spectra of fireballs which produced the L chondrite Kindberg and the aubrite Ribbeck (2024 BX1), showing identified Li I line in both spectra but with notably higher intensity in the aubrite fireball.

\begin{figure}
\centerline{\includegraphics[width=.95\columnwidth,angle=0]{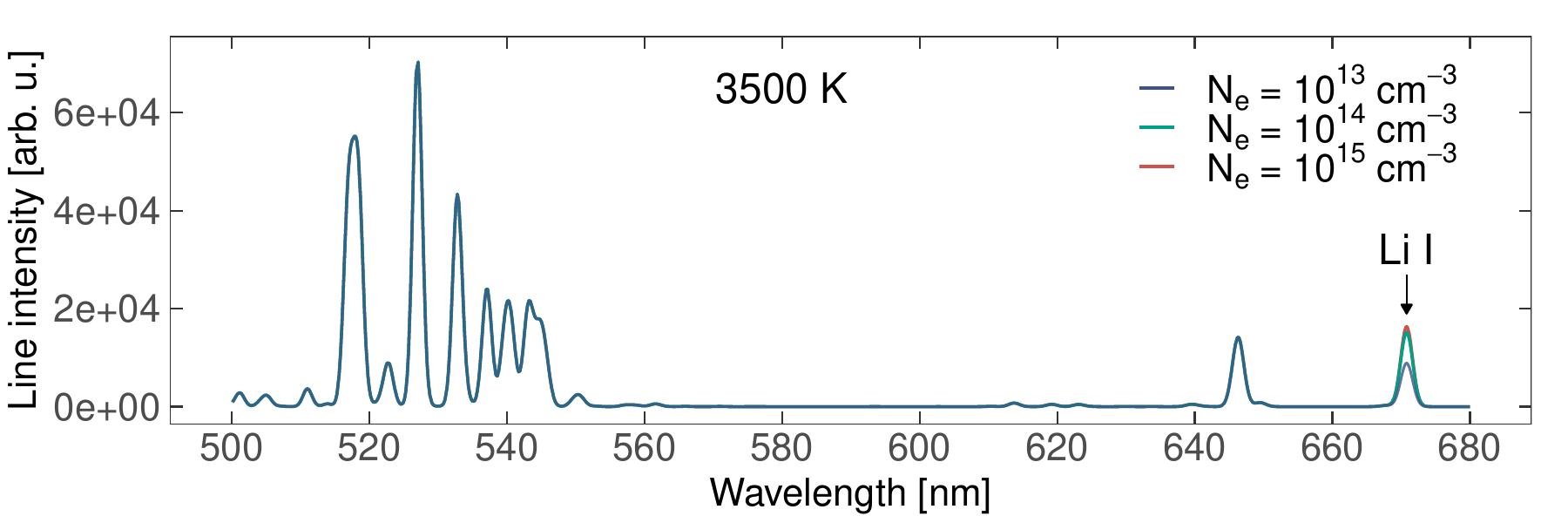}}
\centerline{\includegraphics[width=.95\columnwidth,angle=0]{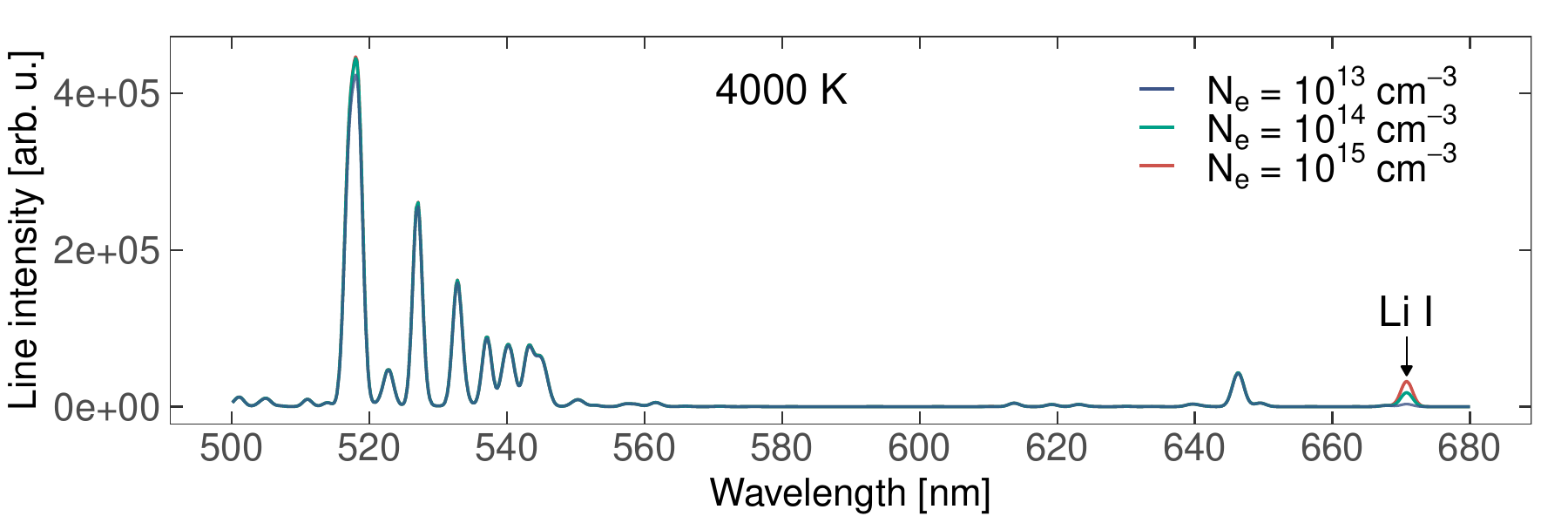}}
\centerline{\includegraphics[width=.95\columnwidth,angle=0]{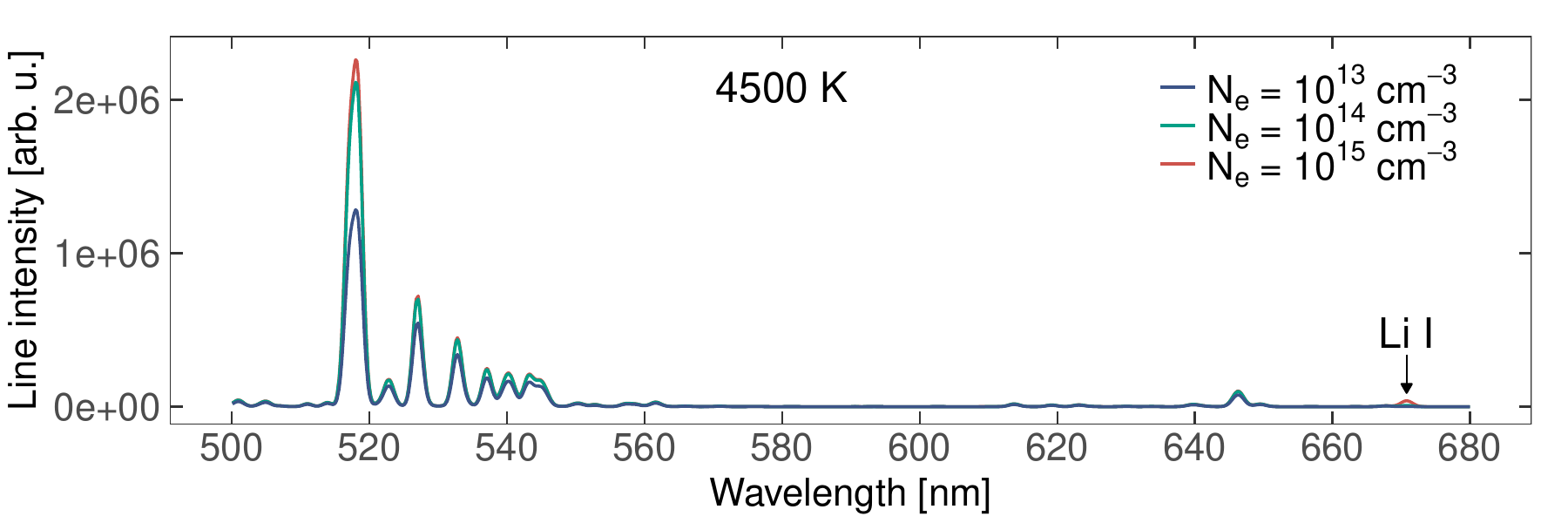}}
\caption[f1]{Model of spectral emission in the 500 -- 680 nm region. The model illustrates conditions at which the detection of the Li I line is possible. The radiative conditions assume optically thin plasma and thermal equilibrium at various temperatures and electron densities. The modeled plasma composition consists of equal parts Mg and Fe with an admixture of 10 ppm of Li.}
\label{Li_model}
\end{figure}

The Li line was first detected in meteors by \citet{1995Boro_Lithium}, and later reported in the bright Benešov bolide \citep{1996Icar..121..484B}. Besides these works, Li has not been commonly studied in meteor spectra analyses. To probe the estimated strength of the Li I line at different radiating conditions, we have performed spectrum modeling assuming optically thin radiating gas in thermal equilibrium at various temperatures and with different Li concentrations (Fig. \ref{Li_model}). We have found that for Li content of 2 ppm (closer to ordinary and carbonaceous chondrites), the Li I line is relatively faint compared to other commonly detected spectral features. With sufficient signal, it can only be recognized at lower plasma temperatures (< 4000 K) and assuming electron densities N\textsubscript{e} > 10\textsuperscript{15} cm\textsuperscript{-3}. The line becomes progressively weaker relative to other features with increasing temperature and decreasing N\textsubscript{e}. 

If we consider a source with a higher Li concentration (10 ppm), closer to the contents found in HED, Martian, or lunar meteorites, the Li I line can be easily identified at temperatures below 4000 K (Fig. \ref{Li_model}). However, at temperatures of 4500 K (often assumed to be a characteristic temperature of the main spectral component in meteors) or higher, straightforward detection of Li I would require N\textsubscript{e} > 10\textsuperscript{15} cm\textsuperscript{-3} and presumably a strong signal of the recorded spectrum. These results suggest that for meteoroids with increased Li concentrations, similar to some of the tested achondrites, and at favorable plasma conditions, the Li I line should be detectable in their spectra and may be used to constrain their composition type.

\subsubsection{Ni I}

\begin{figure}
\centerline{\includegraphics[width=.98\columnwidth,angle=0]{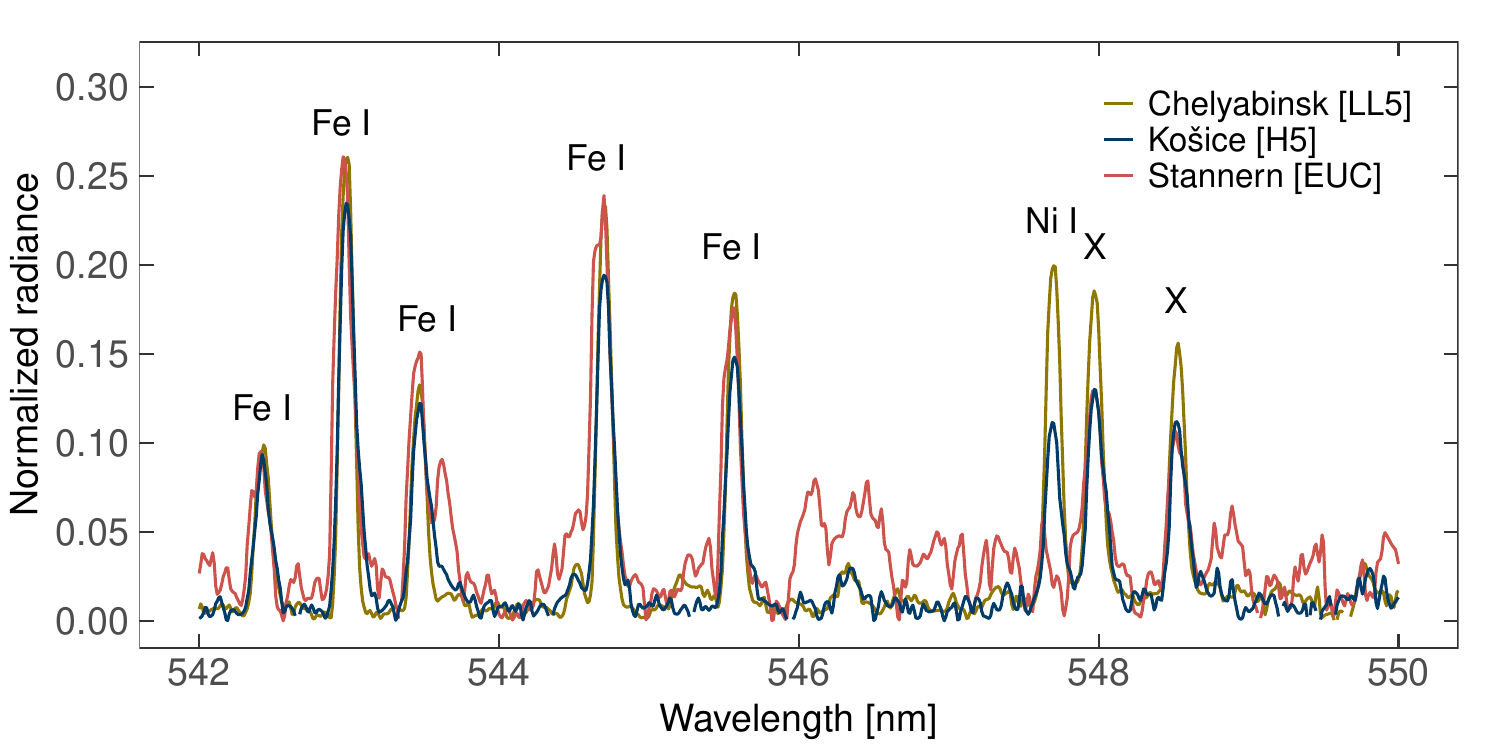}}
\centerline{\includegraphics[width=.98\columnwidth,angle=0]{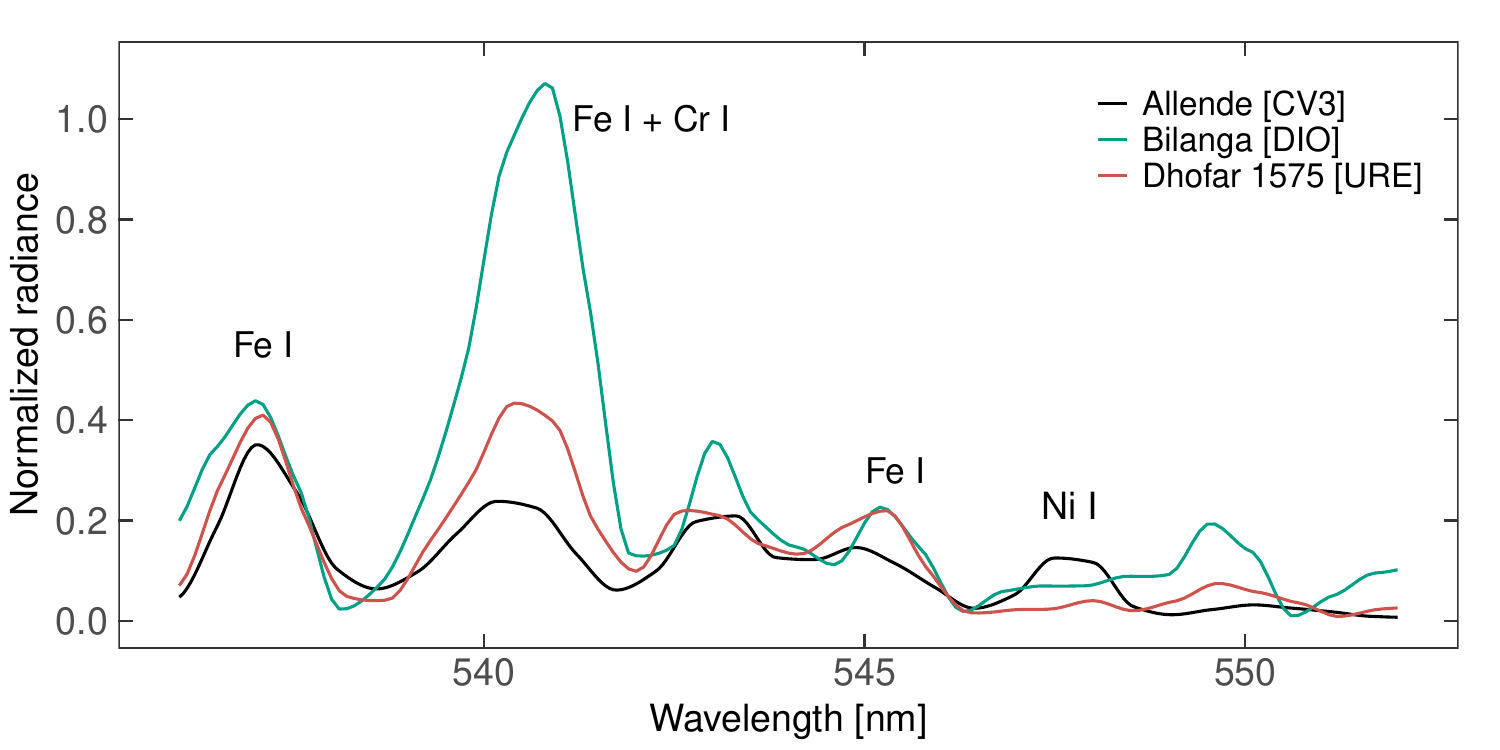}}
\caption[f1]{Comparison of meteorite spectra in the 540 -- 550 nm range demonstrating the variable strength of the Ni I-59 line. Upper panel: Differences between the high-resolution Echelle spectra of three meteorites. The lines marked by X are artifacts of the Echelle reduction. Lower panel: AMOS spectra of three different meteorites illustrating the possible identification of the Ni I in lower-resolution data. The Ni I line was detected in the iron and stony-iron meteorites, and in all chondrites, while it is absent in all tested achondrites (also see Fig. \ref{NiMg}). In both images, the spectra are normalized to unity at 526.95 nm (peak of the strongest Fe I-15 line).}
\label{Ni_profile}
\end{figure}

Most chondrites contain relatively similar bulk content of Ni ($\approx$ 1 -- 1.8 wt$\%$) but its concentration in achondrites is low (Table \ref{T1}). The detection of Ni may pose a challenge for low signal and low-resolution meteor data, as Ni I lines are typically faint and often positioned near strong Fe I lines in visible range spectra. Several Ni I lines were recognized in the analyzed meteorite data. The strongest detected Ni I lines included multiplets of Ni I-59 (547.69 nm), Ni I-32 (385.83 nm), Ni I-33 (377.56 and 380.72 nm), or Ni I-143 (503.54 and 508.08 nm). Strong Ni I lines can be found at wavelengths below 360 nm but were not investigated in detail as this is a region of low spectral sensitivity for most meteor spectrographs. We further focused on measuring the Ni I-59 multiplet, as it is the strongest Ni line in the visible range at relevant meteoric emission conditions. Its measurement needs to be performed carefully in lower-resolution data due to the proximity of the Fe I lines near 545.56 nm and 549.75 nm (Fig. \ref{Ni_profile}). Reliable measurement of Ni emission in similar data would require a relevant model of Fe I emission at given conditions.

\begin{figure}
\centerline{\includegraphics[width=.95\columnwidth,angle=0]{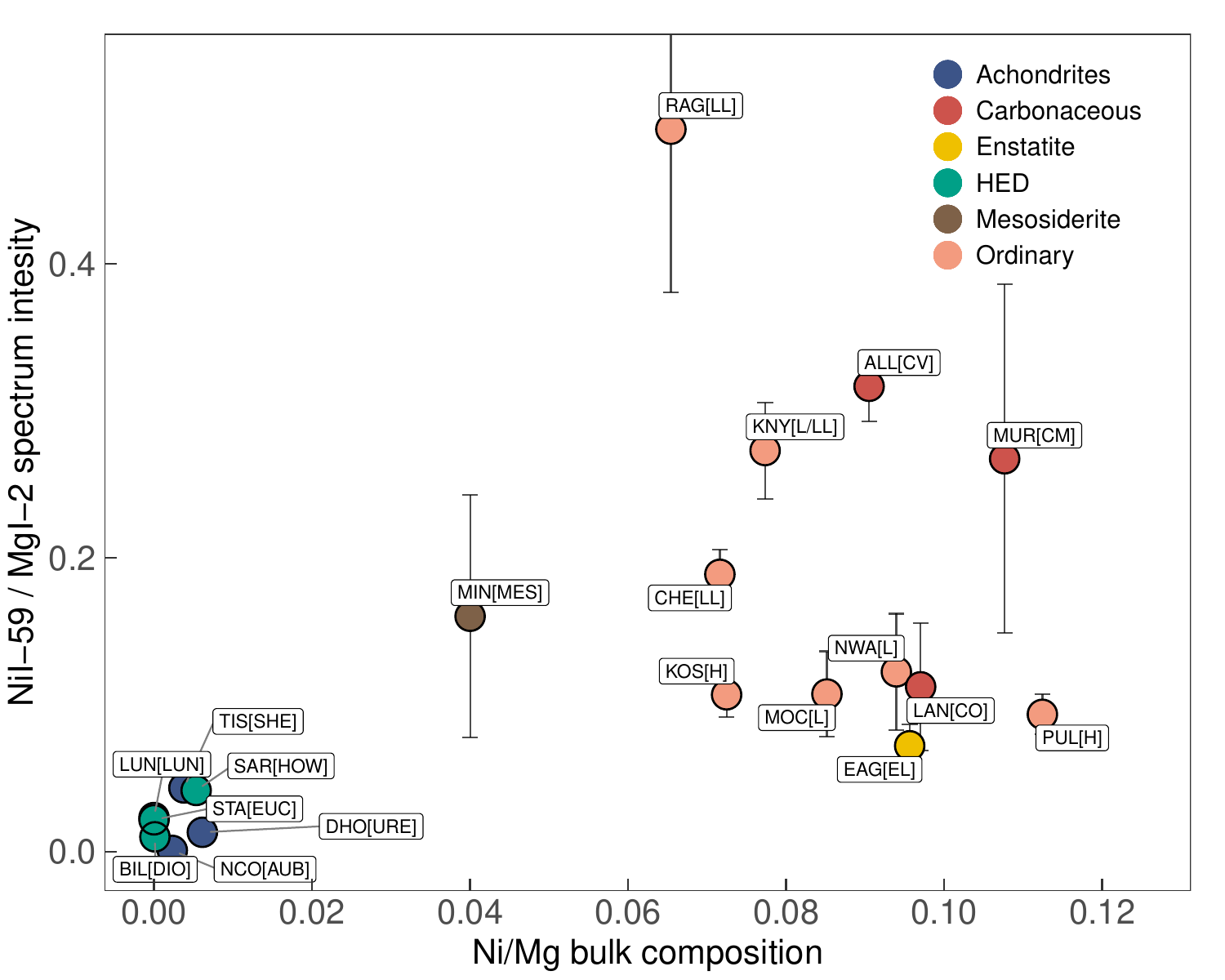}}
\caption[f1]{Ni I-59 / Mg I-2 intensities from high-resolution spectra of meteorite samples. The measured spectral intensities are compared with the original Ni/Mg bulk composition ratio for the same meteorite cases from the literature (references provided in Table \ref{T1}).}
\label{NiMg}
\end{figure}

The results of our measurements of the Ni/Mg intensity ratio in all meteorites, compared with their published bulk chemical composition ratios, are displayed in Fig. \ref{NiMg}. The results from spectral measurements reflect the lack of Ni in achondrites, in which no Ni I lines were identified. On the other hand, Ni I lines have been detected in the iron meteorite, mesosiderite, and in all chondrites. Minor differences in the estimated original Ni/Mg bulk content among the chondrites were not manifested in the measured Ni I-59 / Mg I-2 intensities (Fig. \ref{NiMg}). The detection of Ni I lines in meteor spectra can however be used for a simple distinction between chondritic and achondritic meteoroid types.

\subsubsection{Na I and K I} \label{NaKtext}

As noted in Section \ref{sectionPrevious}, a spectral classification based on the relative intensities of Fe I-15, Mg I-2, and Na I-1 is often applied to meteor spectra analyses. Sodium is an interesting component of meteoroids, as it is universally detected in meteors and can be used to trace the volatile content in meteoroids. The depletion of Na in meteor spectra was linked with the thermal alteration of meteoroids, even causing changes in their structural properties \citep{2019A&A...621A..68V,2022A&A...667A.158B}.

The intensity of the low-excitation Na I-1 doublet depends on plasma temperature and is favored in slower meteors. Since the ablation experiment simulates the conditions of a slow meteor, most studied meteorite spectra contained notably stronger Na I-1 compared to Mg I-2 and Fe I-15 (Fig. \ref{ech_vs_amos}). Na I-1 was the strongest feature in the visible range for most meteorites, except for the iron meteorite sample. In some cases, the Na I-1 doublet was saturated, preventing accurate measurement of its intensity. For meteor spectra studies, the measurement of the Na I-1 line is still of interest but should always be interpreted considering the meteor speed/temperature \citep{2020Icar..34713817M}.

\begin{figure}
\centerline{\includegraphics[width=\columnwidth,angle=0]{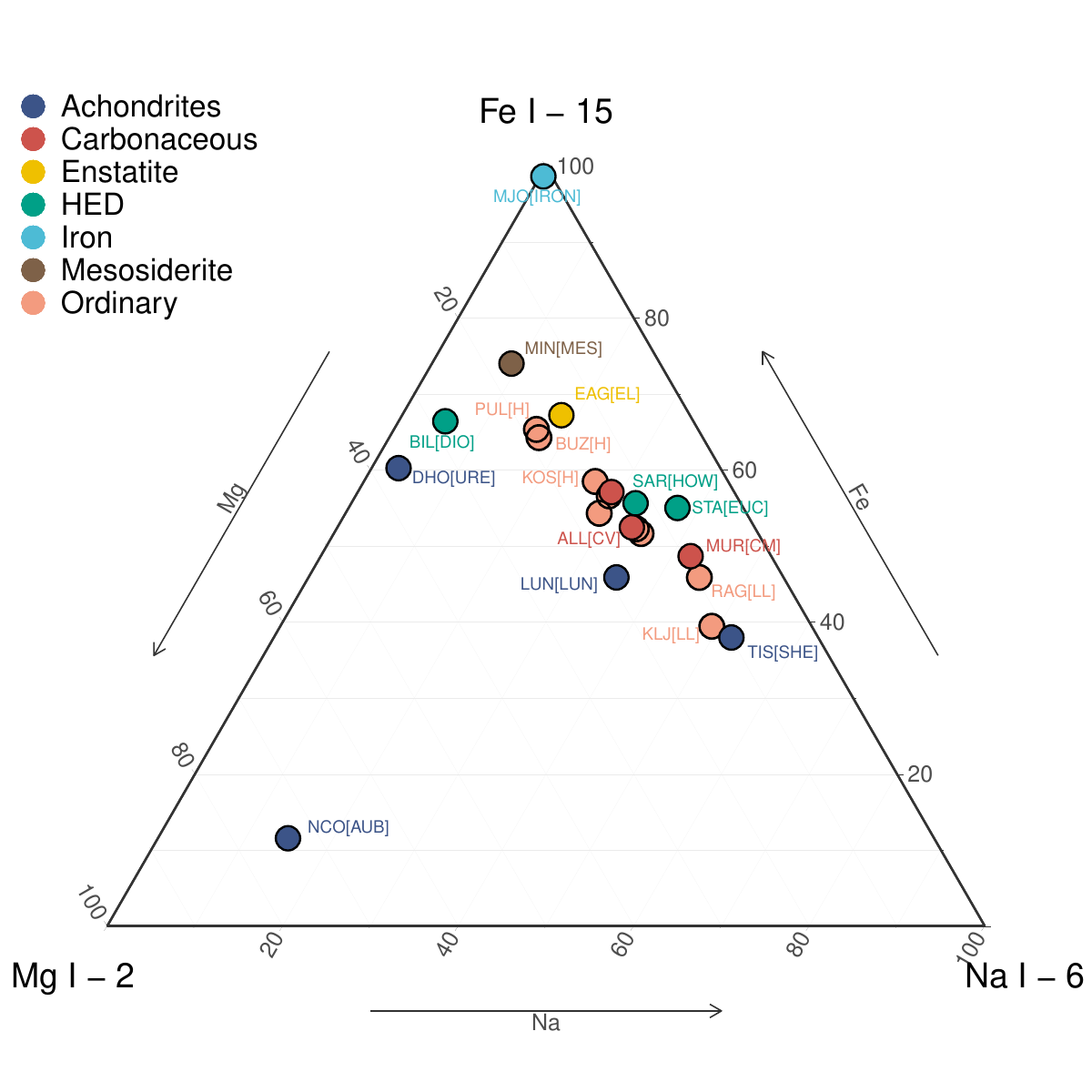}}
\caption[f1]{Measured relative intensities of Mg I-2, Fe I-15, and Na I-6 in spectra of all ablated meteorite samples.}
\label{ternaryMgNaFe}
\end{figure}

\begin{figure}
\centerline{\includegraphics[width=\columnwidth,angle=0]{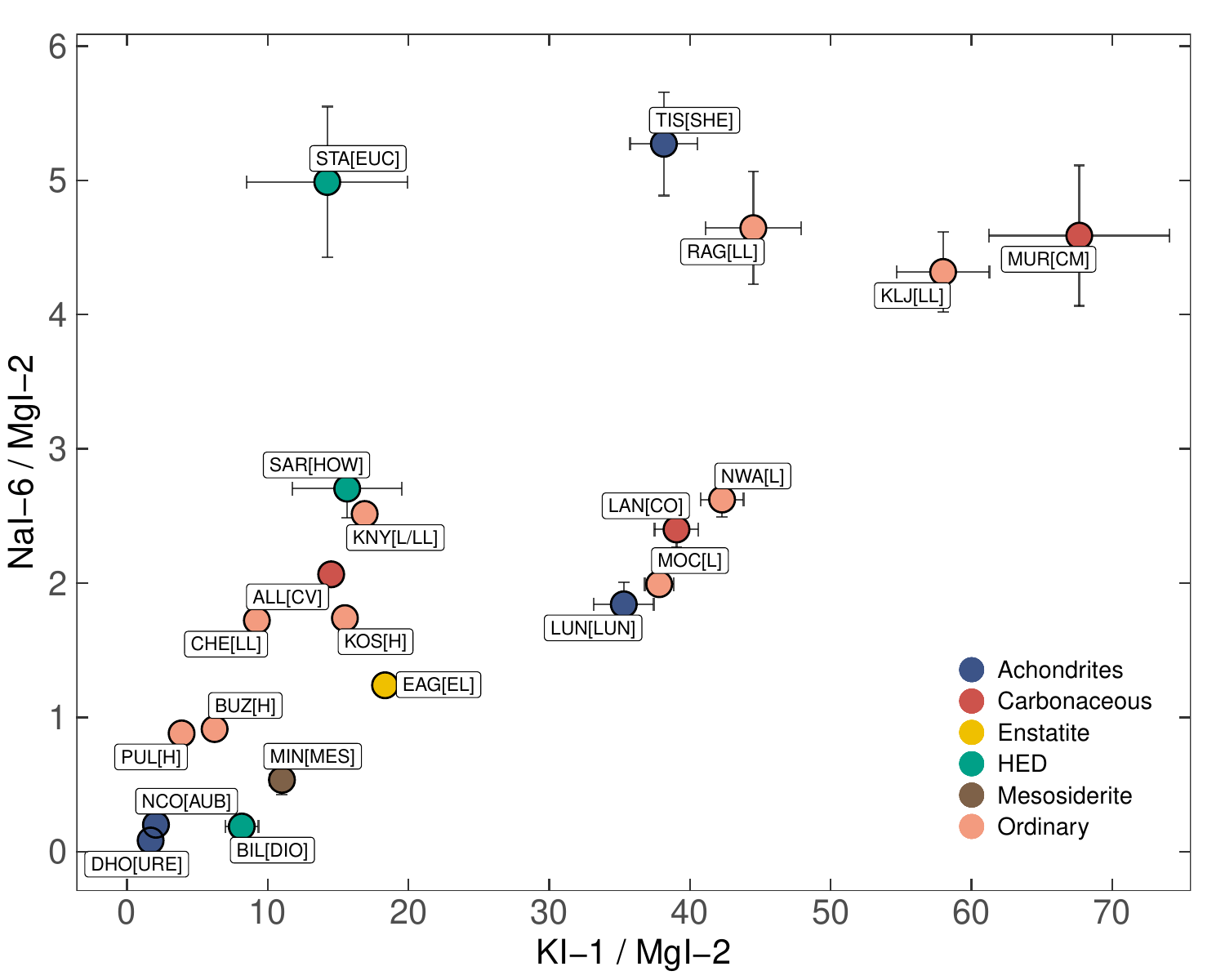}}
\centerline{\includegraphics[width=\columnwidth,angle=0]{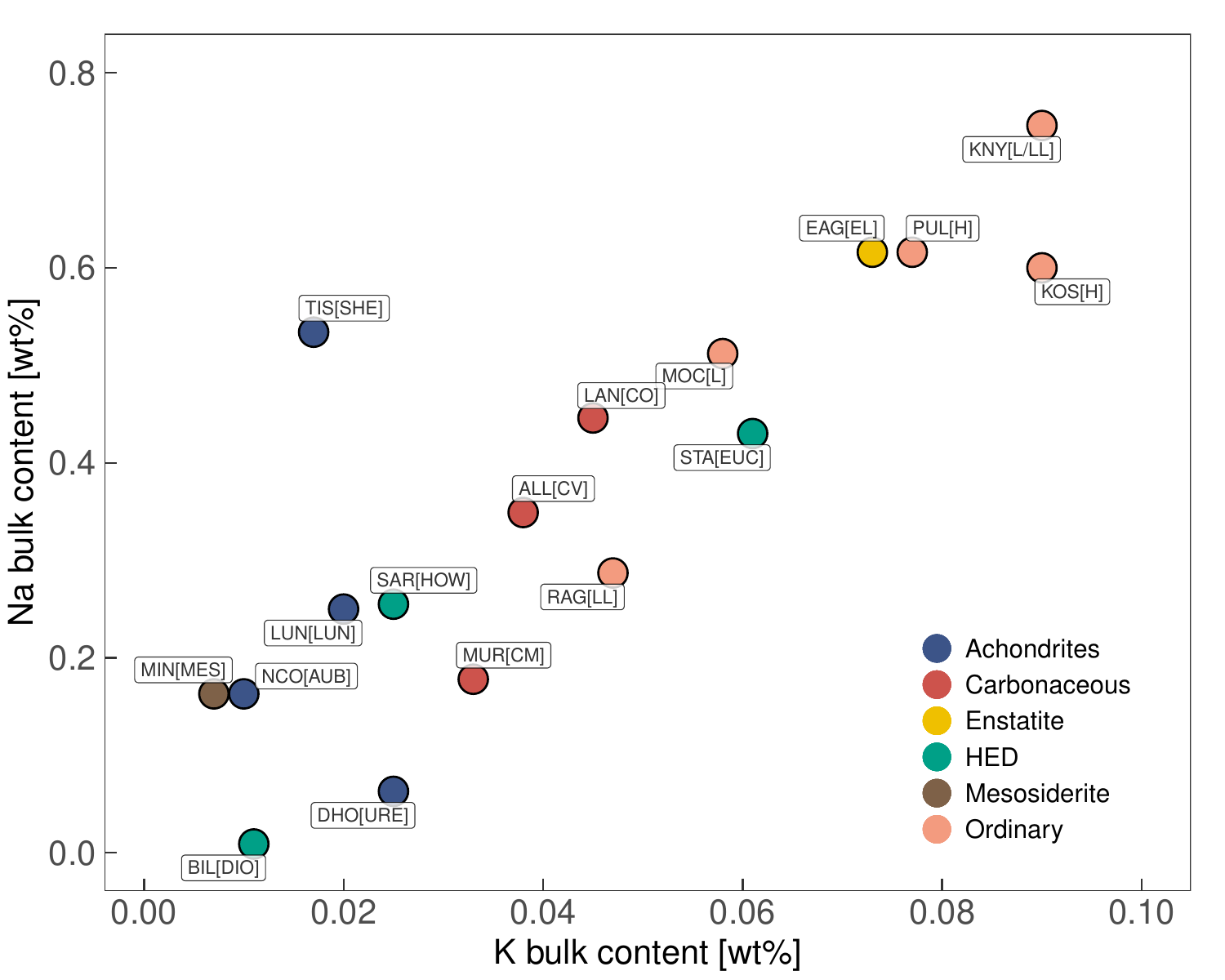}}
\caption[f1]{Measured spectral intensities of Na I-6 and K I-1 relative to Mg I-2 (upper panel) compared with the original bulk content of Na and K (lower panel) in the tested meteorite samples. The Na I-6 lines were not observed in the iron meteorite.}
\label{NaKbulk}
\end{figure}

In our analysis of the meteorite spectra, we instead measured the Na I-6 lines near 568.26 -- 568.82 nm, which are much fainter than Na I-1 but can still be recognized in meteors. A ternary plot analogous to the spectral classification often used for meteor analyses but using Na I-6 intensity instead of Na I-1 is displayed in Fig. \ref{ternaryMgNaFe}. Distinct positions of the individual groups of achondrites can be recognized, as well as the mesosiderite and iron meteorite sample. In this metric, the chondritic meteorites show similar relative Mg intensities. LL chondrites exhibit lower Fe and stronger Na intensity, while H and enstatite chondrites show stronger Fe emission and fainter Na, and L and carbonaceous chondrites are positioned in between these two groups. On average, carbonaceous chondrites (particularly the volatile-rich CO and CM meteorite samples) displayed somewhat enhanced Na I and K I intensities over most ordinary chondrites and achondrites (Fig. \ref{NaKbulk}). Within our dataset, the faintest Na I emission was observed from the iron, ureilite, diogenite, aubrite, and mesosiderite. These results are consistent with their assumed original chemical composition (Table \ref{T1}).

We have also found that the strength of Na I emission generally correlates with the intensity of K I. The observed spectra reflect the strong correlation between the original bulk concentrations of Na and K in meteorites (Fig. \ref{NaKbulk}), caused by the varying feldspar contents in meteorites. The K I-1 lines at 766.49 and 769.90 nm are commonly detected in meteor spectra. Similarly to the Na I-1 doublet, the K I-1 lines can be particularly prominent in slower meteors due to the low excitation energy and high probability of this transition. This finding is supported by the results of our modeling. When we considered a K concentration of 0.5 wt\%, a sharp increase of the K I line intensity was observed at lower temperatures (< 4700 K) and N\textsubscript{e} $\geq$ 10\textsuperscript{15} cm\textsuperscript{-3}. Similar conditions were likely achieved in the plasma wind tunnel experiment, which resulted in strong, often saturated K I lines. The expected electron densities for meteoric plasma in the atmosphere are typically estimated to be one or two orders lower, depending on the speed and size of the meteoroid. This might explain why K I lines are usually less prominent in meteor spectra.

\subsubsection{H I and CN} \label{HCNtext}

At visible wavelengths, H I and CN present the best spectral features allowing us to trace the presence of water and organic compounds in meteoroids. We have presented an analysis of H and CN emission among the ablated meteorites and meteors in separate studies \citep{2022MNRAS.513.3982M, 2023Icar..40415682P}. Here we briefly summarize the main findings concerning their use for constraining specific meteoroid composition classes. 

Though of high scientific interest, the detection of CN in meteor data has been thus far unsuccessful, likely due to the low resolution and sensitivity of most currently employed meteor spectrographs \citep[][and references therein]{1998EM&P...82...71R, 2004AsBio...4...67J, 2023Icar..40415682P}. \citet{2016Icar..278..248B} investigated the presence of CN in the high-quality spectrum of the Benešov fireball. CN was not detected possibly due to the low carbon content of the meteoroid. The B → X $\Delta\nu$ = 0 emission band of the radical CN peaking near 388.33 nm was for the first time clearly recognized in our meteorite spectra dataset. 

\begin{figure}
\centerline{\includegraphics[width=\columnwidth,angle=0]{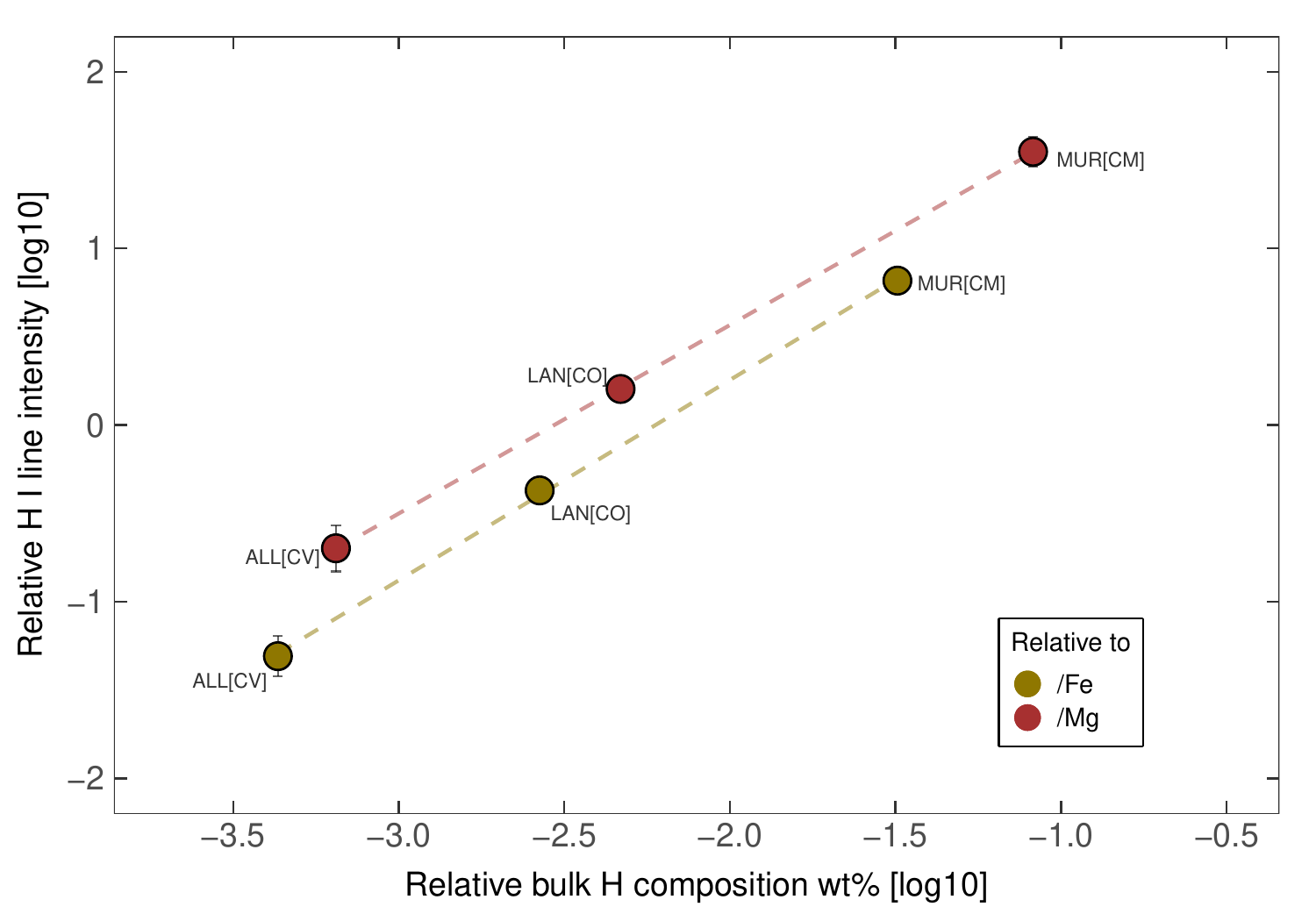}}
\centerline{\includegraphics[width=\columnwidth,angle=0]{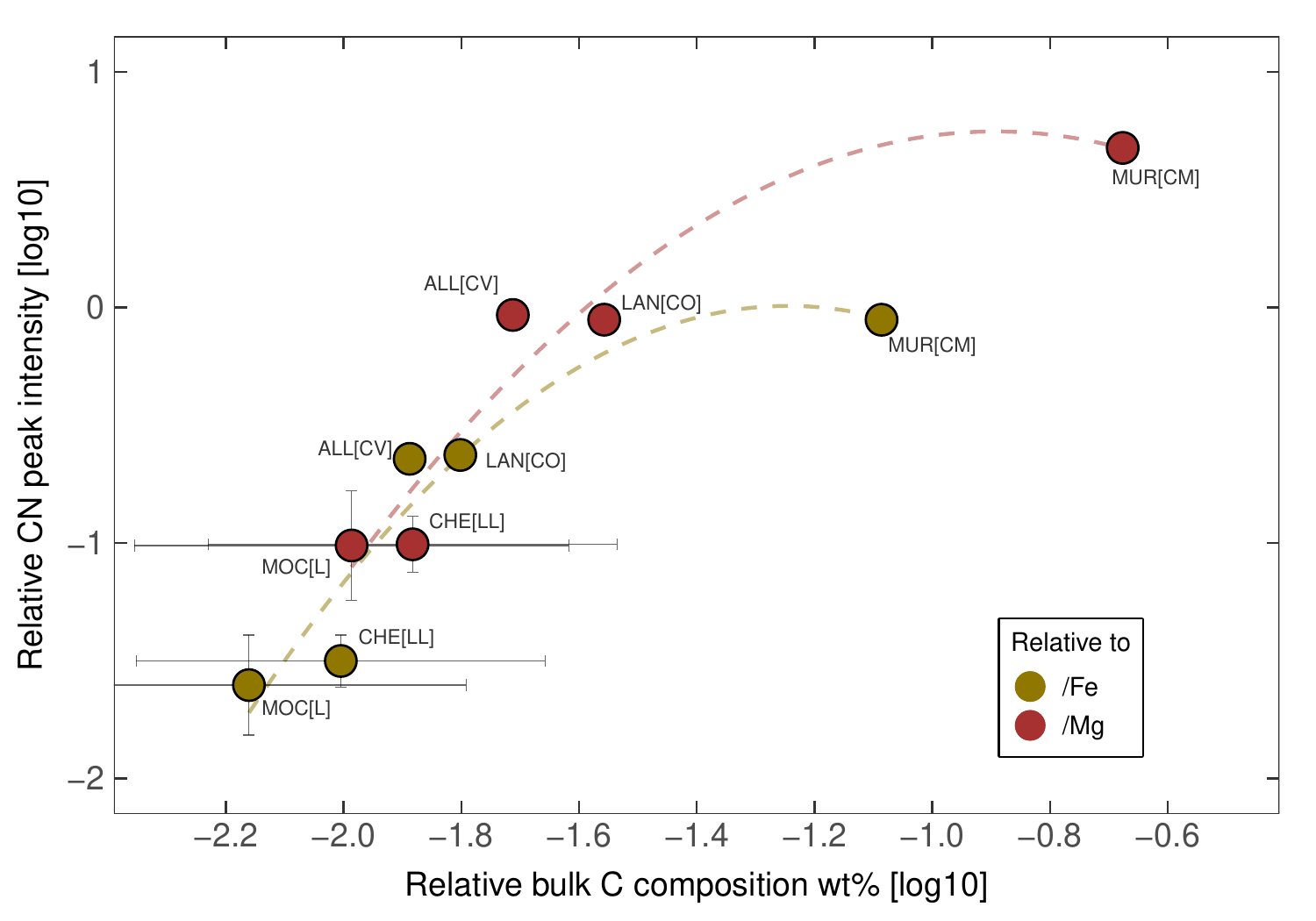}}
\caption[f1]{Measured intensities of the H$\alpha$ line (upper panel) and CN band peak (lower panel) relative to Mg I-2 and Fe I-15 in spectra of tested meteorites compared with their original bulk chemical abundances of H and C. The bulk content of H and C for carbonaceous chondrites is taken from \citet{1985GeCoA..49.1707K}. The average C bulk content and its standard deviation for L and LL ordinary chondrites from \citet{1990Metic..25..323J} was used for the Mocs and Chelyabinsk meteorites displayed in the lower plot. The trend of the correlation between the spectral intensity ratios and meteorite composition is outlined by dashed lines.}
\label{HComplot}
\end{figure}

\begin{figure}
\centerline{\includegraphics[width=\columnwidth,angle=0]{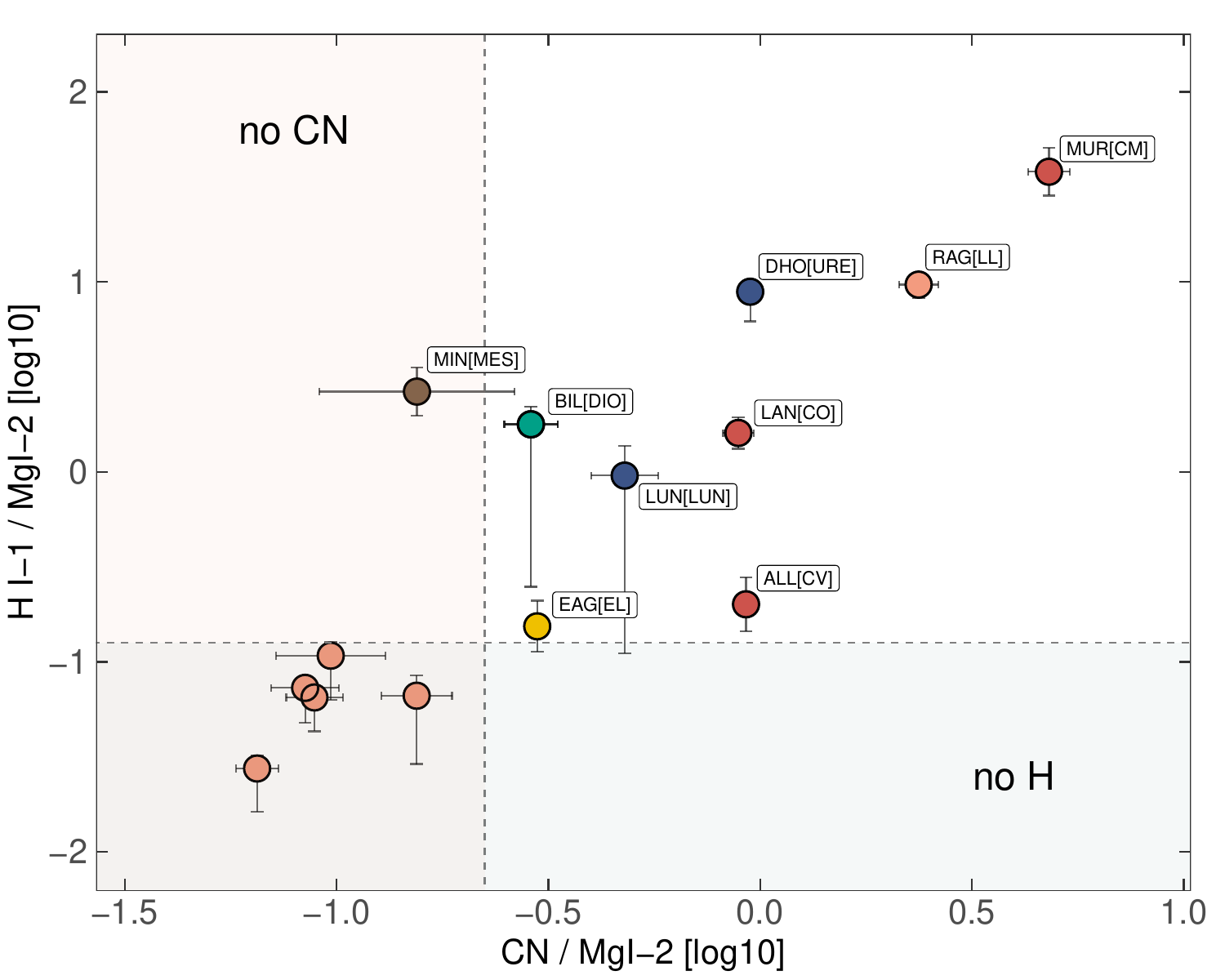}}
\caption[f1]{Measured intensities of H I-1 and CN band peak relative to Mg I-2 in spectra of tested meteorites. H line intensity in meteorites BIL, LUN, and EAG may be affected by a contaminant source of H. Regions characteristic for meteorite spectra with no confirmed H or CN emission are marked in color. The meteorite group color coding is the same as in other figures in this work.}
\label{HCN}
\end{figure}

The H$\alpha$ line emission has been reported in fireball spectra for a long time. It is assumed to originate from a high-temperature spectral component (T $\approx$ 10 000 K) characteristic for the front surface plasma formed in fast and bright fireballs. The source of the hydrogen atoms is assumed to be predominantly in the meteoroid composition, not from the Earth's atmosphere \citep{2004AsBio...4..123J}. This hypothesis is supported by the results from our laboratory ablation experiments \citep{2022MNRAS.513.3982M}. We have found the intensities of both H and CN emissions to be proportional to the bulk contents of H and C within the original meteorite (Fig. \ref{HComplot}). Furthermore, we have found a correlation between the H$\alpha$ line and CN band intensities (Fig. \ref{HCN}) \citep{2023Icar..40415682P}.

Our analysis of H and CN emission in meteorite spectra confirms that these features can be used to trace the contents of water molecules from hydrated minerals and organic compounds in meteoroids. The strongest H and CN emission was observed from the organic-rich CM chondrite Murchison. Both these features were also observed from the other tested carbonaceous chondrites Allende (CV3) and Lancé (CO3.5) with intensity corresponding to the assumed original carbon and hydrogen content (Fig. \ref{HComplot}). Among carbonaceous chondrites, the weakest H$\alpha$ line was observed from the Allende meteorite containing < 0.5 wt$\%$ water, less than for many known ordinary chondrites \citep{1990Metic..25..323J, 2018GeCoA.239...17B}. Relatively strong H and CN emissions were also seen from the carbon-rich ureilite. Moderate to faint H and CN emissions were observed from the lunar meteorite, diogenite, and enstatite chondrite respectively. The measured H I intensity in the diogenite and lunar meteorite is considered uncertain, due to the additional contaminant source of H recognized in the pure plasma flow before the start of meteorite ablation (discussed in \citet{2023Icar..40415682P}).

Faint H and possibly very faint CN were also detected in the aubrite Norton County, which is not displayed in Fig. \ref{HCN} due to its strong Mg I emission. The mesosiderite Mincy exhibited a relatively strong H$\alpha$ emission but no CN. This would suggest a source of H from hydrated minerals such as iron hydroxides, sulfates, or phyllosilicates, rather than C-bound organic compounds. Finally, strong H and CN emissions were observed from the LL chondrite Ragland, known to contain high water content ($\approx$ 2.45 wt$\%$). We assume the Ragland spectral features are partially affected by the moderate to high terrestrial weathering of the meteorite \citep{1986Metic..21..217R}. Other analyzed meteorites, including all tested ordinary chondrites, did not exhibit apparent H or CN emission. For more discussion on the H and CN contents in the studied meteorites, please refer to \citet{2022MNRAS.513.3982M} and \citet{2023Icar..40415682P}.

Overall, H and CN present useful diagnostic tools for recognizing the presence of hydrated minerals and organic compounds in meteoroids. Strong H and CN may indicate classes analogous to primitive carbonaceous chondrites (most notably CM and CI groups), specific meteorites with higher C content (such as ureilites), or samples with higher water content. Our data suggests that most ordinary chondrites will not display H and CN emissions. While some ordinary chondrites were found to contain > 2 wt$\%$ water content (particularly the LL group), this is usually characteristic of meteorite finds affected by terrestrial weathering. Fresh ordinary chondrites from meteorite falls were found to have generally low water content \citep{1990Metic..25..323J}, so the same can be assumed for the corresponding ordinary chondrite-like meteoroids detected in the atmosphere. 

For meteor spectrographs with low sensitivity in the red wavelength region, the H$\beta$ line can be measured instead of the H$\alpha$. In such cases, a strong signal from the meteor is required, as the H$\beta$ was approximately 10 times fainter than H$\alpha$ in our meteorite ablation data. Accurate measurement of the CN band intensity is challenging for the currently employed meteor spectrographs. Our findings \citep{2023Icar..40415682P} suggest that more efficient detection of CN in meteors may be achieved by implementing sensitive higher-resolution spectrographs and focusing on meteor emission in the early stages of the ablation in the upper atmosphere, as supported by the arguments of \citet{2010Icar..210..150B}.
 
\subsection{Other minor species} \label{minor}

In addition to the previously discussed diagnostic species, we have identified a few other minor species in meteorite spectra. These may also provide benefits for constraining meteoroid composition but are generally faint, making their identification challenging in lower-quality meteor data. Here we give an overview of the strongest identified lines of these species, for a potential search of minor elements in rich meteor spectra.

Copper was detected in all meteorites except for the eucrite Stannern. The strongest Cu I lines were observed at 324.76, 327.39, 510.55, 515.32, and 521.82 nm. Carbonaceous chondrites exhibited higher Cu I line intensities compared to ordinary chondrites. Several lines of cobalt were also identified, with the strongest Co I peaks at 345.23, 350.23, 350.63, 384.55, and 389.41 nm. These faint lines were more notable in chondrites and the mesosiderite, while they were not reliably identified in achondrites. Vanadium was detected in several meteorites. Va I lines appear strongest in the diogenite Bilanga and shergottite Tissint but were also likely identified in several ordinary and carbonaceous chondrites. The best-resolved lines of Va I were found at 437.92, 440.84, and 459.14 nm.

We have additionally searched for strontium (Sr I and Sr II), barium (Ba I and Ba II), sulfur (S I), iridium (Ir I), and platinum (Pt I) but did not reliably detect any spectral lines for these elements. Considering the collected spectra are exceptionally rich, the presence of other species not covered in this work is plausible and will be investigated further.

\subsection{Undetected species} \label{undetected}

Here we provide a brief discussion focused on spectral features that were not detected in the studied meteorite spectra, but can be identified in meteor observations and may be useful for constraining the meteoroid composition type. 

No aluminum lines were detected in any of the studied meteorites. Al I emission would present a potentially suitable trace feature, given the increased Al content in lunar samples, eucrites, howardites, shergottites, and even carbonaceous chondrites over ordinary chondrites. The best candidate for detecting aluminum is the low excitation Al I-1 line near 396.15 nm, previously reported in fireball spectra \citep{1994P&SS...42..145B}. According to our modeling, at LTE conditions and a range of plasma temperatures relevant for the wind tunnel ablation (3500 -- 6500 K), the line should dominate in its wavelength region for meteorites with Al content higher than 2 wt$\%$. For example, the tested lunar meteorite has an estimated 15 wt$\%$ Al content (Table \ref{T1}), but no Al emission was detected. Furthermore, emission bands of AlO were reported in spectra of space debris and some fireballs \citep{2016Icar..278..248B} but were not visible in our experiment. These results imply that Al I atoms were underabundant in the radiating plasma.

A similar trend in compositional differences between the studied meteorite types as observed for aluminum is also typical for calcium. This is due to the common occurrence of clinopyroxene and plagioclase minerals in specific non-chondritic meteorite classes. In meteor spectra, Ca is mainly represented by the strong Ca I-2 line near 422.64 nm and the Ca II-1 lines near 393.37 nm and 396.80 nm. None of these features were observed in any of the analyzed meteorites. Again, we have conducted spectrum modeling assuming a sample with a higher concentration of Ca (3 wt$\%$) at a wide range of temperatures, and electron densities relevant to the plasma wind tunnel experiment. The modeling predicts that at such Ca concentration (relevant for several of our samples), the Ca I and Ca II lines should be prevalent in their spectral region at most conditions. Even considering the increased optical thickness of these lines, the complete lack of Ca I and Ca II cannot be explained only by the physical parameters of the plasma. 

Finally, our modeling also predicts that several lines of Ti I should be visible in the spectra of most ablated meteorites. Several Ti I can be recognized in meteor spectra, particularly in the 450 -- 510 nm range (e.g., lines of Ti I-42 peaking near 453.3 nm or Ti I-37 near 498.2 nm). Similar to Al I and Ca I, no Ti I lines were confidently detected in our data. On average, increased Ti contents (over values typical for ordinary and carbonaceous chondrites) are expected in eucrites, howardites, shergottites, and some mesosiderites \citep{1990Metic..25..323J}. 

The lack of Al I, Ca I, and Ti I lines in the studied spectra suggests incomplete evaporation of Al, Ca, and Ti during the meteorite ablation. This is also supported by previous fireball spectra analyses, where these refractory elements were underabundant in the radiating gas \citep{1993A&A...279..627B}. Our results confirm the effect of incomplete evaporation and suggest that under laboratory conditions, the evaporation of refractory elements was even less efficient, as no lines of these species were identified. Even considering their underabundance in meteor plasma, studying the variations of Ca I, Al I, and Ti I line intensities and their correlation with other identified features may help recognize meteoroids with a composition similar to lunar meteorites, eucrites, howardites, shergottites, or mesosiderites. On the other hand, carbonaceous and ordinary chondrites will likely exhibit comparable relative intensity ratios of these species. We note that recently, lines of Al I, Ca I, and Ti I were detected in the spectrum of the fireball that produced the enstatite-rich aubrite Ribbeck (2024 BX1) \citep{2024A&A...686A..67S}. The intensities of these lines in the aubrite spectrum were stronger than in the compared spectrum of an L-type chondrite Kindberg.

\section{Meteorite type characteristics} \label{charcteristics}

Based on the results presented in Section \ref{results}, here we summarize the identified spectral characteristics of the analyzed meteorite types. A simplified overview of the distinct spectral features characteristic for each type is given in Table \ref{T2}. The features are divided into strong (lines/bands with above-average intensity within our dataset), weak (lines/bands with below-average intensity), and missing (lines/bands not detected at all). Note that this designation is meant to be interpreted as relative to the average spectral intensities of each species within the dataset. Here, the assumed average intensities are usually characteristic of the most common meteorite types -- ordinary chondrites (and for most criteria also valid for carbonaceous chondrites). It can be assumed that the statistical distribution of material types within the meteoroids in interplanetary space will differ from the distribution of meteorite types on Earth. Therefore, the reference 'average' spectral line intensities may be shifted from our sample (also assuming a wider range of physical conditions of the radiating plasma). The guide provided in Table \ref{T2} can be used for interpreting the relative spectral differences between meteoroids.

\begin{table*}[]
\small\begin{center}
\caption{Summary of the identified distinct spectral features of different meteorite types.}
\label{T2}
\vspace{0.1cm}
\resizebox{\textwidth}{!}{
\begin{tabular}{llllll}
\hline\\[-6pt]
\multicolumn{1}{l}{\multirow{2}{*}{\textbf{Meteorite type}}} & 
\multicolumn{1}{l}{\multirow{2}{*}{\textbf{Reference sample}}} & 
\multicolumn{3}{c}{\textbf{Distinct spectral features}} & 
\multirow{2}{*}{\textbf{Remarks}} \\ [3pt]
\multicolumn{1}{c}{} & & \textbf{Strong} & \textbf{Weak}   &  \textbf{Missing}  &  \\ [3pt]
\hline \\[-4pt]

\multirow{2}{*}{H and L ordinary} & Košice (H5), Pultusk (H5), Buzzard Coulee (H4),  & \multirow{2}{*}{}  & \multirow{2}{*}{Cr, Li} & \multirow{2}{*}{H, CN} & compact Fe, Mn, Si, Mg  
\\ &  NWA 869 (L3-6), Mocs (L5-6) & & & & \\ [5pt]

\multirow{2}{*}{LL ordinary} & Knyahinya (L/LL5), Ragland (LL3.4), & \multirow{2}{*}{often Cr, Na, K} & \multirow{2}{*}{Li} & \multirow{2}{*}{H, CN} & variable group; 
\\ & Chelyabinsk (LL5), Kheneg Ljouâd (LL5/6) & & & & compact Fe, Mn, Si, Mg \\ [5pt]

CM carbonaceous & Murchison (CM2) & H, CN, Na, K & Cr, Mn, Si &  & Fe, Mg similar to OC \\ [5pt]
CO carbonaceous & Lancé (CO3.5) & H, CN  & Cr, Mn, Si, Li &  & Fe, Mg similar to OC \\ [5pt]
CV carbonaceous & Allende (CV3) & CN & Cr, Mn, Si, Li &  & Fe, Mg similar to OC \\ [5pt]
Enstatite & Eagle (EL6) & Mn & Cr, Li & Ni &  Fe, Mg, Si similar to OC\\ [5pt]
Eucrite & Stannern & Si, Cr, Li, Al*, Ca* & Mg, K & Ni, H, CN &  \\ [5pt]
Diogenite & Bilanga & Cr, Si, Mg & Fe, Na, K & Ni & moderate CN in our sample \\ [5pt]
Howardite & Sariçiçek & Si, Cr, Al*, Ca* & Mg, K & Ni, H, CN &  \\ [5pt]
Martian   & Tissint (shergottite) & Si, Mn, Cr, Al*, Ca* & Mg & Ni &  \\ [5pt]
Lunar & NWA 11303 & Si, Al*, Ca* & Fe, Mg, Cr & Ni &  moderate H, CN in our sample \\ [5pt]
Aubrite & Norton County & Mg, Si & Fe & Ni & faint H, CN in our sample \\ [5pt]
Ureilite & Dhofar 1575 & CN, Mg, H, Li & Si, Na, K, Mn & Ni &  \\ [5pt]
Mesosiderite & Mincy & Fe, Cr, Si & Na, K & CN & moderate H in our sample \\ [5pt]
Iron & Mount Joy & Fe & Li, Cr, Mn & H, CN, Si, Mg &  \\ [4pt]
\hline \\[-20pt]
\end{tabular}}
\end{center}
\tablefoot{This guide is based on our analysis of the referenced meteorite samples ablated at conditions corresponding to the ablation of a low-velocity asteroidal meteoroid. The asterisk denotes species not detected at meteorite ablation conditions but expected to be prominent in meteor spectra.}
\end{table*}

\subsection{Ordinary and enstatite chondrites} \label{Ordinary_discussion}

\begin{figure}
\centerline{\includegraphics[width=\columnwidth,angle=0]{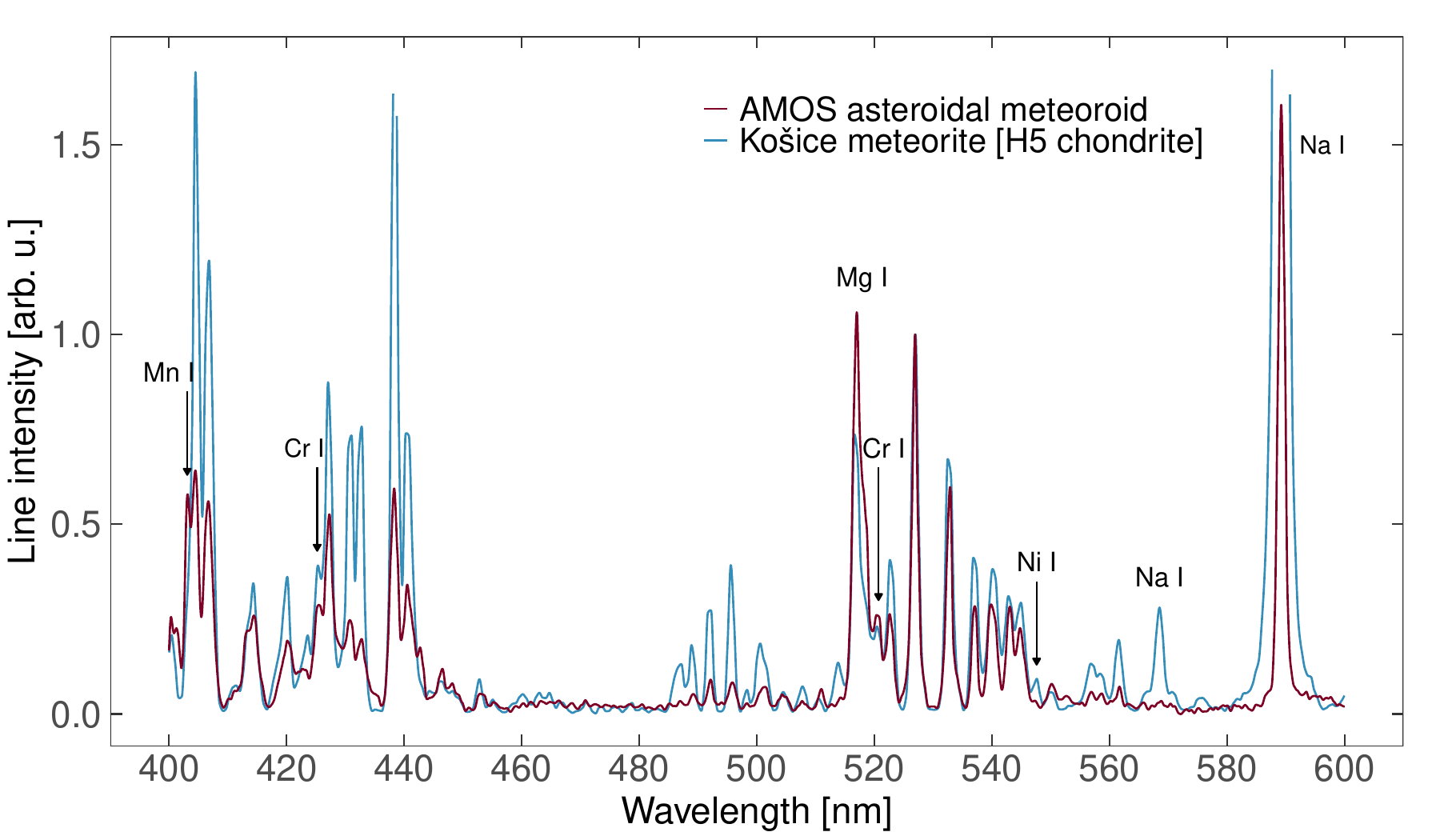}}
\caption[f1]{Emission spectra of the ordinary chondrite Košice captured by AMOS during the ablation experiments and an asteroidal meteor captured by the AMOS network above Australia. Spectra are normalized to unity at 526.95 nm (Fe I-15 line). Most unmarked emission lines belong to Fe I.
}
\label{AMOScomparison}
\end{figure}

Ordinary chondrites are the most common meteorites found on Earth and were also the most represented meteorite type in our dataset. The observed spectra are in many ways similar to spectra of normal-type asteroidal meteors (Fig. \ref{AMOScomparison}). The main differences between the spectra displayed in Fig. \ref{AMOScomparison} are the increased intensities of low excitation lines in the laboratory spectrum, and the overall lower signal from the meteor resulting in fainter lines in the blue wavelength region.

Spectra of ordinary chondrites usually contained lines of Fe I, Mg I, Na I, Mn I, Cr I, K I, Si I, Ni I, and Li I. The Mg I / Fe I intensity ratios are comparable to most asteroidal meteor spectra, with somewhat fainter Mg I lines, probably due to lower plasma temperature and incomplete evaporation of Mg. The entire group can be constrained by similar Mn I, Si I, Mg I, and Fe I intensities (Fig. \ref{SiMg_MnMg} and \ref{SiFe_MnFe}). H and L chondrites also showed similar Cr I and Li I intensities and no emission of H or CN. The LL group appears to be spectrally more heterogeneous than other chondrites, making its classification based on spectra challenging. The heterogeneity of LL chondrites is not surprising, given that many meteorites classified into this group are breccias and the group lacks a well-defined hiatus in bulk composition. While their spectra are similar in some aspects to H and L chondrites (Fe I, Mg I, Mn I, and Si I), some LL chondrites show notably higher intensities of Cr I, Na I, or K I. We have also detected stronger Ni I emission from all LL chondrites compared to H and L samples (Fig. \ref{NiMg}), which was not indicated by the assumed comparable original bulk contents of Ni in these meteorites (Table \ref{T1}).

The Ragland LL chondrite is the most spectrally distinct from the other analyzed ordinary chondrites, exhibiting strong intensities of Li, Cr, Ni, Na, K, H, and CN. Ragland is a meteorite find with reported moderate to high weathering and some atypical compositional properties resulting from its brecciated nature \citep{1986Metic..21..217R}. Note that significant enhancement in the average H\textsubscript{2}O and subtle enhancement in the average C bulk concentrations are observed in weathered ordinary chondrites (finds) compared to fresh ordinary chondrites (falls) \citep{1990Metic..25..323J}. We believe the atypical spectral signature of Ragland is partly affected by the reported terrestrial weathering of this meteorite and partly stems from the brecciated nature of the interchondrule material, which likely occurred during its accretion \citep{1986Metic..21..217R}. 

The only analyzed enstatite meteorite in our dataset was the EL6 chondrite Eagle. In most aspects, it was spectrally similar to ordinary chondrites. In comparison, it showed a slightly increased Fe I / Mg I intensity ratio and weaker emission of Ni I. The main distinction from the other studied meteorite classes was in its enhanced Mn I emission (Fig. \ref{SiMg_MnMg}) and the detection of faint H and CN emission (Fig. \ref{HCN}) reflecting slight overabundance of C and H\textsubscript{2}O bulk content over most of the studied ordinary chondrites \citep{1990Metic..25..323J}. More enstatite meteorite samples are required to characterize the dispersion of spectral properties of this group. The available limited analyses suggest that most enstatite chondrites have similar bulk composition in most aspects, but can include samples with enhanced total Fe content and lower Mg content \citep{1990Metic..25..323J,2022GeCoA.318...19P}.

\subsection{Carbonaceous chondrites}

Within our experiments, we have tested and analyzed the spectra of three different carbonaceous chondrites. The observed intensities of Fe I, Mg I, and Cr I in carbonaceous chondrites were similar to ordinary chondrites. The carbonaceous samples can, however, be distinguished by fainter Si I and Mn I intensities (e.g., by measuring Si I-3 and Mn I-2 intensities relative to Mg I-2 and Fe I-15 as in Figs. \ref{SiMg_MnMg} and \ref{SiFe_MnFe}), and by the presence of the tracers of hydrated minerals and organic compounds -- H I and CN (Fig. \ref{HCN}). Similarly to ordinary and enstatite chondrites, all carbonaceous chondrites showed emission of Ni I. In two samples (Allende and Murchison), the measured relative intensity of Ni I / Mg I was enhanced over the H and L chondrites in our dataset. On average, carbonaceous chondrites (particularly CM and CO) showed somewhat increased Na I and K I intensities over ordinary chondrites, and most other achondrites (Fig. \ref{NaKbulk}).

The three groups of carbonaceous chondrites were distinct in their relative intensities of H and CN, corresponding with the estimated original contents of water and organic compounds (Fig. \ref{HComplot}). The strongest H and CN intensities were detected from the CM meteorite Murchison, moderate intensities from the CO meteorite Lancé, and weaker H and moderate CN were detected from the CV meteorite Allende. While we only studied one meteorite per carbonaceous chondrite group, their specific bulk contents of H and C are representative of meteorites with the same classification \citep{1985GeCoA..49.1707K, 1990Metic..25..323J}. 

\subsection{HEDs}

The basaltic HED meteorites were found to be spectrally distinctive from the more common chondritic classes as well as from other types of achondrites. One of the main differences is the characteristically increased Cr I intensities (Fig. \ref{bulk_c} and \ref{CrMgFe}). All HEDs also showed stronger Si I emission and fainter Fe I emission than observed in chondrites. Unlike chondrites, all HED emission spectra lacked the presence of Ni I lines but contained relatively strong Li I. Mn/Fe and Si/Fe intensity ratios can be useful for distinguishing HEDs from spectra of other types of achondrites (Fig. \ref{SiFe_MnFe}). 

Several spectral differences were also revealed between the individual eucrite, diogenite, and howardite meteorites. As expected, the biggest differences were observed between the eucrite and diogenite samples. Howardites are formed as regolith breccia of eucrites and diogenite materials \citep{PAPIKE_1998_achondrites}. This corresponds with the observed spectral properties, as they usually display line strengths between those measured in eucrites and diogenites. The most notable spectral and compositional difference between the studied eucrite and diogenite meteorites is in the concentration and corresponding relative intensity of Mg (Table \ref{T1}, Fig. \ref{bulk_a}). Since many of the diagnostic spectral features studied in this work were compared relative to Mg I, this causes a seemingly wide dispersion of the individual HED meteorites in many of the presented scatter plots. Moreover, the eucrite shows fainter Cr I and stronger Na I lines, while the diogenite was depleted in Na. All studied HEDs also showed relatively faint K I intensities (Fig. \ref{NaKbulk}). However, this finding is inconclusive due to the lower quality of the eucrite and howardite spectra in the far-red wavelength region. 

The only HED meteorite with detected CN and potentially faint H emission was the diogenite Bilanga. The original bulk contents of H and C in the studied HEDs are unknown. There is evidence that the HED parent asteroid Vesta contains water, with reported D/H values suggesting that the volatile material on Vesta accreted from carbonaceous chondrites \citep{2019GeCoA.266..568S, 2016M&PS...51.1110B}. More HED samples should be investigated to characterize the expected occurrence of volatile spectral features in HED meteoroids. Based on our results and the available HED chemical composition analyses, we do not expect the H and CN to be characteristic spectral signs of HEDs.

As discussed in Section \ref{undetected}, Ca I, Al I, and Ti I were not detected in our data. Given the distinct bulk contents of Al, Ca, and Ti in HEDs (Table \ref{T1}, \citet{PAPIKE_1998_achondrites}), we expect that their intensities will be convenient diagnostic spectral features of this group, with strong emission expected from the eucrites and howardites, and missing or weak Ca I, Al I, and Ti I lines in diogenite spectra. One potential diogenite meteor spectrum with properties corresponding to our findings (strong Mg, Si, Cr, faint Na, Fe) was previously reported by \citet{1994ASPC...63..186B}.

\subsection{Other achondrites}

\begin{figure*}
\centerline{\includegraphics[width=.9\textwidth,angle=0]{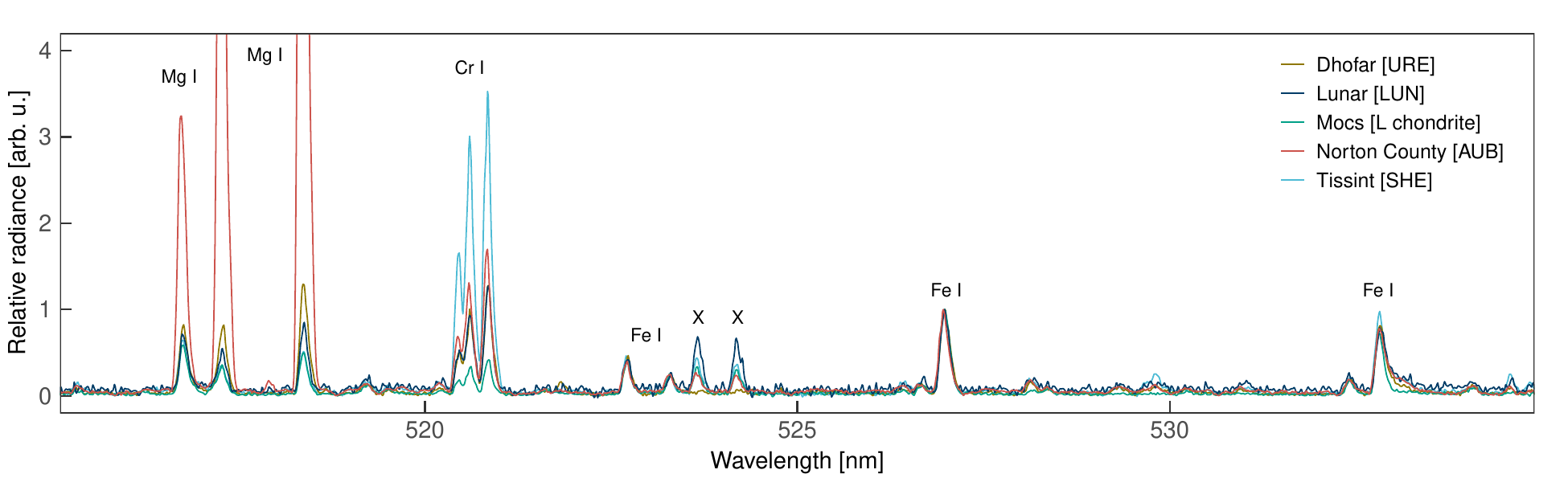}}
\caption[f1]{Emission spectra of the aubrite, lunar, Martian, and ureilite achondrites and the ordinary chondrite Mocs in the 515 -- 535 nm region. Spectra are normalized to unity at 526.95 nm (Fe I-15 line). The Mg I-2 lines in the aubrite spectrum peak at a relative radiance of around 11.5 arb. u. The lines marked by X are artifacts of the Echelle reduction and do not represent meteorite spectral lines.}
\label{Achondrite_comp}
\end{figure*}

Besides HEDs, we have analyzed the spectra of four less common achondrite groups -- Martian (shergottite), lunar, aubrite, and ureilite. Each of the representative meteorites has a distinct composition (Table \ref{T1}) corresponding to various observed spectral signatures (Table \ref{T2}). Among the few common spectral features of these classes is the lack of Ni I lines and Li I intensities higher than found in chondrites. Except for the shergottite Tissint, these achondrites also show fainter Fe/Mg intensities than chondrites. Comparison between the spectra of these achondrites and an ordinary chondrite in the 515 -- 535 nm region is displayed in Fig. \ref{Achondrite_comp}

\subsubsection{Aubrite}

\begin{figure*}
\centerline{\includegraphics[width=.9\textwidth,angle=0]{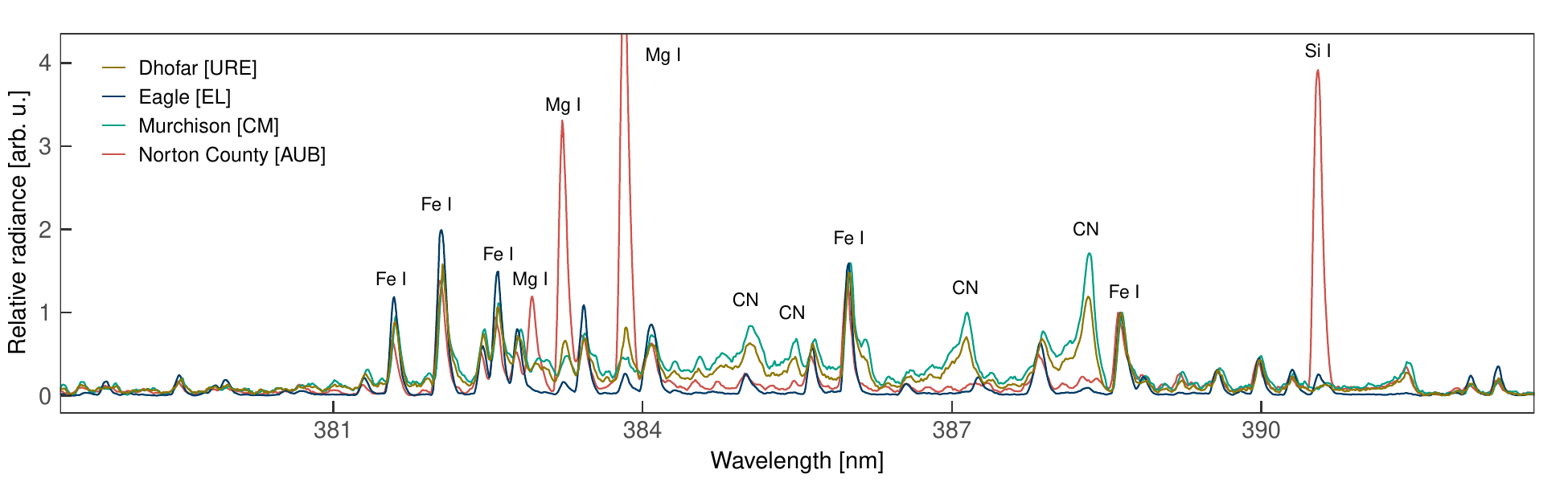}}
\caption[f1]{Emission spectra of the ureilite, carbonaceous, aubrite, and enstatite chondrite meteorites in the 380 -- 392 nm region. Main emission features are highlighted. Most unmarked lines belong to Fe I. In this region, the ureilite displays enhanced CN and Mg I emission, the CM chondrite strong CN, the aubrite very strong Si I, Mg I, and low Fe I, and the enstatite chondrite mainly Fe I lines, similarly to ordinary chondrites. Spectra are normalized to unity at 388.63 nm (Fe I-4 line).}
\label{380-390nm}
\end{figure*}

Aubrites are primitive, often brecciated, enstatite achondrites of atypical mineralogy. Their Mg-silicate-rich composition is well represented by the analyzed meteorite Norton County, which exhibited very distinct Mg- and Si-rich, and Fe- and Cr-poor emission spectrum (Figs. \ref{bulk} and \ref{CrMgFe}). The Norton County spectrum also included moderate intensity Na I and K I lines. A faint H I emission was observed from the ablating aubrite. Bulk H\textsubscript{2}O content of 0.3 $\pm$ 0.2 $\%$ was reported in the Norton County meteorite \citep{2020Sci...369.1110P}, though recent analyses suggest that this H\textsubscript{2}O content may be predominantly caused by terrestrial weathering \citep{2023E&PSL.62018341P}. We have also detected faint CN emission in the early stages of the aubrite ablation \citep{2023Icar..40415682P}, which gradually decreased with time. Given the source of C predominantly in the outer layers of the meteorite and the low intensity of the CN band, we consider the occurrence of CN in the aubrite spectra inconclusive. The comparison of the spectra of aubrite, ureilite, carbonaceous chondrite, and enstatite chondrite in the 380 -- 392 nm region is displayed in Fig. \ref{380-390nm}.

Recently, a meteorite classified as an aubrite (named Ribbeck) was recovered after the impact of a small asteroid 2024 BX1. The spectrum of the corresponding fireball was captured by the European Fireball Network systems \citep{2024A&A...686A..67S}. The preliminary analysis of the aubrite fireball spectrum properties is in agreement with our spectral analysis of the aubrite Norton County. The emission spectrum from the Ribbeck fireball is poor in Fe with more prominent Mg, Na, Li, and Ca lines. Ti and Al intensities stronger than in the compared spectrum of the L-chondrite Kindberg were also detected \citep{2024A&A...686A..67S}. The Si emission was not analyzed in their work.

\subsubsection{Lunar and Martian meteorites}

The lunar meteorite NWA 11303 also displays a Fe-poor and Si-rich emission spectrum, but unlike the aubrite, it contains faint Mg I lines reflecting the typical Mg-poor composition of feldspathic lunar meteorites \citep{Papike1998_lunar, 2021M&PS...56..206K}. The distinct Si-rich and Mg-poor spectra allow straightforward identification of the lunar meteorite within our dataset (Fig. \ref{SiMg_MnMg}). Based on the mean bulk chemical composition of known lunar and Martian meteorites, Al I and Ca I lines in meteor spectra will likely present convenient diagnostic features. CN emission was only observed in the early stages of the ablation of the lunar meteorite. H I line was also observed in the lunar spectrum, but its intensity is uncertain due to the additional contaminant source of H in this experiment (Fig. \ref{HCN}). We assume that the contribution of H atoms from the lunar meteorite in the observed radiating plasma was low.

Like the lunar meteorite, the spectrum of the Martian shergottite Tissint shows strong Si and fainter Mg intensities compared to most other meteorites. The two spectra can be however easily distinguished. The Martian meteorite exhibits stronger lines of Mn I, Cr I, Fe I, and Na I (Fig. \ref{CrMgFe} and \ref{SiFe_MnFe}). The Na I emission was among the strongest within our dataset (Fig. \ref{NaKbulk}). The shergottite Tissint contained the largest bulk content of Fe among the studied achondrites (Table \ref{T1}). Since this sample also contains less Mg, the measured Fe/Mg intensity ratio appears comparable to ordinary chondrites (Fig. \ref{bulk_a}). While both lunar and Martian meteorite spectra showed low Mg I intensities, they were notably fainter in the lunar spectrum. Similarly to other achondrites, both lunar and Martian meteorite spectra lacked Ni I lines and presented relatively strong Li I. No CN or H emission was detected in the Martian meteorite spectrum.

\subsubsection{Ureilite}

One ureilite -- Dhofar 1575 was tested during our laboratory ablation campaigns. Ureilites are a rare type of achondrites with both igneous and primitive characteristics and contain a high percentage of carbon in the form of graphite and nanodiamonds \citep{1992Metic..27..327G}. We did not find any reference for the bulk composition of the tested Dhofar 1575. The observed spectral properties of this ureilite were compared with the mean bulk composition computed from six ureilites analyzed in \citet{1970SSRv...10..483V}. The ureilite spectrum can be recognized by notable CN, H, and Mg emissions. The CN and H intensities were the strongest among all studied achondrites (Fig. \ref{HCN}). The detected Mn/Mg, and Si/Mg intensity ratios were among the lowest in our entire dataset (Fig. \ref{SiMg_MnMg}). Compared to ordinary and carbonaceous chondrites, the ureilite shows lower Fe/Mg intensity (Fig. \ref{CrMgFe}). The spectrum was also characterized by low Na I and K I intensities (Fig. \ref{NaKbulk}). Similarly to other achondrites, a relatively strong Li I line and no Ni I were detected. 

We note that Dhofar 1575 is a meteorite found in 2009 with unknown terrestrial age, but reported low weathering degree \citep{Dhofar1575}. \citet{1970SSRv...10..483V} has previously reported lower H content (0.13 $\%$) in an unoxidized ureilite, where H enters the composition of hydrocarbons. Higher H contents ($\approx$ 0.40 $\%$) detected in two oxidized ureilites were linked to terrestrial weathering. Given the low weathering of Dhofar 1575, we may assume that the detected strong H and CN emissions are intrinsic features of the original ureilite parent meteoroid. 

\subsection{Iron and stony-iron meteorites}

Identifying iron meteoroids is usually straightforward even from lower-resolution spectra. Fe I lines dominate the visible wavelength range with other common species absent or faint. FeO emission can be recognized, as was reported from meteor observations \citep[][and references therein]{2021MNRAS.500.4296P}, though it was not apparent under the laboratory ablation conditions. Our data from the ablation of the iron meteorite Mount Joy suggests that Ni I lines and faint emission of other species such as Cr I, Mn I, or Na I may be recognized with sufficient signal and spectral resolution. This is consistent with recent analyses of iron meteor spectra \citep{2019A&A...625A.106C}, in which the detection of weak Ca I and Ca II lines was additionally reported. These species represent the silicate inclusions in iron meteoroids.

One stony-iron meteorite was tested in our ablation campaigns -- the mesosiderite Mincy. The bulk composition of cm-sized stony-iron bodies will strongly depend on the proportion of the metal and silicate components. Mesosiderites, which formed as breccias, were thought to better represent a stony-iron meteoroid ablation, rather than the strongly differentiated pallasites. We did not find any reference for the total bulk Fe content of the mesosiderite Mincy. Average bulk Fe contents of mesosiderites are expected to range between 30 to 70 wt\% with an average value of around 50 wt\% \citep{2004AMR....17..231N}. The observed spectral properties of Mincy, showing Fe/Mg intensity comparable to ordinary chondrites (Fig. \ref{CrMgFe}), suggest that our cm-sized mesosiderite sample had lower metal content. 

Still, strong Fe emission, along with increased intensities of Cr I and Si I were found to be characteristic of the mesosiderite spectrum (Fig. \ref{CrMgFe} and \ref{SiMg_MnMg}). On the other hand, the spectrum showed fainter Na I and K I lines compared to most other meteorites (Fig. \ref{NaKbulk}). The measured Li I and Ni I intensities were distinctly between achondritic and chondritic meteorites (Fig. \ref{LiMg_LiFe} and \ref{NiMg}). Interestingly, we have detected relatively strong H emission from the mesosiderite. Mincy is the only sample in our dataset showing clear H and no CN emission (Fig. \ref{HCN}), suggesting a different source of the radiating H atoms. As we reported in \citet{2023Icar..40415682P}, stronger H$\alpha$ intensity was observed in the early stages of the mesosiderite ablation, which gradually decreased in the second half of the ablation (after approximately 0.35 seconds). This implies a source of H in the outer layer of the meteorite. Mincy is a meteorite find, with a reported low weathering degree and higher porosity compared to other mesosiderites \citep{1998M&PS...33.1231C}. The potential influence of terrestrial weathering on the observed H emission cannot be ruled out. 

\section{Conclusions} \label{sec:conclusions}

Spectral properties and diagnostic features of a wide range of ablated meteorite samples serving as asteroidal meteor analogs were presented. The observed spectral characteristics were found to be consistent with those of low-speed asteroidal meteors. Using spectrum modeling, we estimate the temperatures of the radiating meteorite plasma of 3700 -- 4800 K, similar to the main temperature component in meteors. We demonstrate that the observed line intensity variations generally trace the differences in the original meteorite chemical composition and can be used to constrain the composition type. The characteristic spectral features of the studied meteorite types, based on relative line intensities of Mg I, Fe I, Na I, Cr I, Mn I, Si I, H I, CN, Ni I, and Li I, are summarized in Table \ref{T2}. The measurement and analysis of the presented suite of spectral features in meteor data can improve our abilities to classify meteoroid composition types.

Ordinary chondrite spectra were found to be most similar to typical asteroidal meteor spectra, characterized by comparable intensities of Fe, Mn, Si, Mg, and Ni lines, and a lack of H and CN. Carbonaceous chondrites can be distinguished by the presence of H and CN emission and fainter Mn and Si lines compared to ordinary chondrites. H line was detected in laboratory meteor analogs due to additional excitation, potentially from electron impacts, but may not be recognized in slower meteors where thermal excitation is dominant.

Different types of achondrites show distinct spectral properties but collectively lack Ni lines and exhibit fainter Fe and stronger Li emission compared to chondrites. The basaltic HEDs were characterized by strong Si and Cr lines. The aubrite spectrum is distinctly rich in Mg and Si and poor in Fe, as in agreement with the meteor spectrum of the recent Ribbeck aubrite meteorite fall. The Martian and lunar meteorites both show faint Mg lines and strong Si, but can be distinguished by the faint Fe and Cr emissions from the lunar sample. The spectrum of the carbon-rich ureilite contains strong CN, Mg, and H emissions and faint Si and Na lines. Measurement of Al and Ca lines detectable in meteor data can further improve the distinction of achondritic material types. The stony-iron mesosiderite is characterized by relatively strong Fe, Cr, and Si emissions, and weak intensities of Na and K. Spectra of iron meteors are the easiest to recognize, as they are composed dominantly of Fe lines, with only faint contributions from Ni, Cr, Mn, and Na.

Our results confirm the effect of incomplete evaporation of refractory elements Al, Ti, and Ca, which were not present in the studied meteorite spectra. The underabundance of these elements in the radiating meteor plasma was previously reported by \citet{1993A&A...279..627B}. The presence of other minor species including Co I, Cu I, and V I was confirmed in meteorite spectra.

The spectral characteristics of different meteorite types presented here may be beneficial for interpreting differences within larger datasets of meteor spectra. The analyzed meteorite spectra were all obtained at similar conditions, simulating the entry of a slow meteor. A direct comparison with absolute line intensity ratio values from meteorites is suitable for slow meteors radiating under comparable conditions. Since the conditions for meteor phenomena can vary more significantly, the interpretation of meteor spectra should also consider meteor temperature, brightness, and altitude. Analyzing the intensities of the presented spectral lines in a large dataset of meteor spectra may provide additional insights into the diversity of meteoroid materials in the Solar System, which likely differs from the distribution of meteorite classes found on Earth. Collecting fireball spectra from meteorite-dropping meteoroid impacts can provide a valuable extension of this dataset and validate the presented results. 

\section*{Data availability}

Data sets generated during this study are available from the corresponding author upon reasonable request. 

\begin{acknowledgements}

We thank Dr. Jiří Borovička for a helpful review of this manuscript. We are also grateful to the High Enthalpy Flow Diagnostics Group team of the Institute of Space Systems, University of Stuttgart for the meteorite experiments. The assistance from the technicians at IRS and NHMV workshops is greatly appreciated. This work was supported by ESA under contract no. 4000128930/19/NL/SC and contract no. 4000140012/22/NL/SC/rp, by the Slovak Grant Agency for Science grant VEGA 1/0218/22, and by the Slovak Research and Development Agency grant APVV-16-0148.
     
\end{acknowledgements}

\bibliographystyle{aa}
\bibliography{references}

\end{document}